\documentclass[final,twocolumn,5p,authoryear]{elsarticle}

\usepackage{amssymb}

\usepackage{epsfig}         
\usepackage{color}          
\usepackage{times}          
\usepackage{amsmath}        
\usepackage{graphicx}

\journal{Journal of High Energy Astrophysics}

\begin{document}

\begin{frontmatter}



\title{Jet formation model from accretion disks of electron-ion-photon gas }


\author[label1]{E. Katsadze}
\address[label1]{Department of Physics, Faculty of Exact and Natural Sciences,
     Javakhishvili Tbilisi State University (TSU), Tbilisi 0179, Georgia}
\author[label1]{N. Revazashvili}
\author[label1,label2]{N. L. Shatashvili}
\address[label2]{TSU Andronikashvili Institute of Physics, TSU, Tbilisi 0177, Georgia}

\begin{abstract}
The problem of Astrophysical Jet formation from relativistic accretion
disks through the establishment of relativistic disk-powerful jet equilibrium
structure is studied applying the Beltrami-Bernoulli equilibrium approach of
\citealt{SY2011,yso}. Accretion disk is weakly magnetized consisting of fully
ionized relativistic electron-ion plasma and photon gas strongly coupled
to electrons due to Thompson Scattering.
Analysis is based on the generalized Shakura-Sunyaev $\alpha $-turbulent
dissipation model for local viscosity (being the main source of accretion),
in which the  contributions from both the photon and ion gases are taken
into account. Ignoring the self-gravitation in the disk we constructed
the analytical self-similar solutions for the equilibrium relativistic
disk-jet structure characteristic parameters in the field of gravitating
central compact object for the force-free condition. It is shown, that
the magnetic field energy in the Jet is several orders greater compared
to that of accretion disk, while jet-outflow is locally Super-Alfv\'enic with
local {\it Plasma-beta} $< 1$ near the jet-axis. The derived solutions can be
used to analyze the astrophysical jets observed in binary systems during
the star formation process linking the jet properties with the parameters
of relativistic accretion disks of electron-ion-photon gas.
\end{abstract}


\end{frontmatter}




\section{Introduction}
\label{Intro}

Observations suggest that the star formation process is often accompanied by the
accretion disk (AD), with the significant intrinsic link found
to the strong bipolar/unipolar outflows/jets/winds streaming out from them
carrying away the matter, energy and angular momentum of the accreting matter,
thus effectively promoting the development of a star and playing the important
role in the disk-evolution (\citealt{bland,Ustyugova,Dullemond07}).
When the accretion rate is sub-Eddington and the opacity very high,
the standard thin accretion disk is formed. It is geometrically thin
in the vertical direction and is made of a relatively cold gas,
with a negligible radiation pressure. Properties of AD-jet outflow
systems can be explored from the observations of:
young stellar objects (YSOs, Herbig-Haro (HH) objects, T-Tauri
and Herbig Ae/Be stars) as well as of the galactic objects
(black holes (BH), neutron stars, white dwarfs (WD)), extragalactic
supermassive BHs and their disks.  Thin disk is not always formed,
specifically in the case of BHs Binaries and X-ray Binaries (of which
disk is mostly optically thin while the polar regions of compact object
can be either optically thin or optically thick -- e.g. case of
X-Ray Pulsars, BH-s) (see e.g. (\citealt{Hartmann,Ferreira08,Mirabel})
and references therein).

It is expected, that large-scale outflows met in various
active astrophysical objects being the collimated long-lived
structures related to accreting disks surrounding the central
compact objects [see e.g. (\citealt{Begelman}) and references
therein] are playing an important role in stellar evolution,
including late stages of their lives and their final fates.
The most luminous objects of the Universe (observed in different classes
of narrow binary systems as well as Active Galactic Nuclei (AGNs)
and quasars) are fed by accretion of matter onto central compact objects.
It is believed that many stars are born in the binary systems going
through one or more phases of the mass-exchange (\citealt{winget,DAZ,
tremblay,mukai}). Numerous observed accreting WDs are often
surrounded by an accretion gas of companion star / disk (\citealt{Begelman,mukai}).
Cataclismic Variables (CVs) and Symbiotics are representing
the accreting white dwarf binaries (AWBs) being
important laboratories for accretion and outflow physics.
In nova-like CVs the outflow velocities can vary within 200-5000 km/s
(\citealt{AD_modeling,kafka,Diaz}).

It is well-known that magnetic fields play important roles in accelerating
jets in different ways (\citealt{bland,uchida,shibata3,shibata});
also the gas pressure may become important in some cases
(\citealt{takahara,shibata4}), being often dominant in the disks
specifically for weakly magnetized cases (\citealt{Koide,Widrow}).
Then, the observable properties of jets/outflows as well as the ADs
depend not only on the accretion mechanism in disk but on the jet-acceleration
mechanism as well. Therefore, one has to consider both heating and
acceleration mechanisms in the unified approach in addition to the
formation stage of jets for better comparisons between theoretical
models and observations (\citealt{zanni,shibata4}). At the same time
the main mass/energy source of the ejecta is the disk-flow material/energy
released through the accretion process. Hence, jet velocities are
intrinsically correlated with the accretion rates dependent on the unified
disk-jet system dynamics; the central object dynamics may
play the additional role in the formation of relativistic outflows/jets
[\citealt{Livio}-- powerful jets are produced by systems in which on top of an
accretion disk threaded by a vertical field, there exists an additional source
of energy/wind, possibly associated with the central object (for example,
stellar wind from porotostar may accelerate YSO jets, as estimated by
\citealt{Ferreira1997,Ferreira,Ferreira07}].

AGNs, hosting super-massive accreting black holes, are commonly
(\citealt{Ferreira23,FerreiraUltra}) believed to be powered by an optically thick,
geometrically thin AD that emits in the optical and UV
energy range (\citealt{Shakura1973}); some of which also show hard X-ray
emission that is assumed to result from the disk photons Compton
scattering off thermal electrons in the so-called hot corona,
located somewhere near the BH. According to
\citealt{Ferreira23}, when the local magnetization is high,
the effect of the formed jets on the disk structure can be tremendous
since the jets' torque efficiently extracts the disk angular
momentum, significantly increasing the accretion rate.
As a result such Jet Emitting Disk (JED) has a much lower
density in comparison to the standard accretion disk (SAD)
and produces the hard X-ray emission attributed to the hot corona.
The parameter space for stationary JED solutions correspond to magnetization
in the range [0.1,1]. In the outer region, where the magnetization
is small ($\ll 1$) a SAD (\citealt{Shakura1973}) is present.

As mentioned above, the observable properties of jets, such as the
local velocities, energies, opening angle, mass-loss rate
depend on their formation mechanism, AD characteristic properties
and dynamical acceleration as well as heating mechanisms. Moreover,
we know that morphology, geometry and velocities of the jets can
be used to estimate the mass, luminosity and/or age of the YSOs
(see \citealt{Bally2016} and references therein) and compact objects
(see e.g. \citealt{begelman5,bland4}. Explorations of the jet-outflows (see, e.g.,
\citealt{Ioannidis2012,Smith2014,Zhang2014} for YSOs) provide an unbiased
observational data that can be used to test the theoretical models of
the equilibrium disk-jet structures.  In (\citealt{SY2011}) it
was shown, that there exists a general principle that dictates
a marked similarity in macroscopic accreting disk-jet outflow
geometry despite the huge variety of the scaling parameters such as
Lorentz factor, Reynolds number, Lundquist number, ionization fractions, etc.,
characterizing different systems indicating that magnetic field
(pre-existed or generated) could play crucial role in jet-acceleration
and its collimation, and not for disk-jet structure formation
(see in addition \citealt{Yoshida2012,yso}). For the jet acceleration
the magnetic mechanism (when the global poloidal magnetic fields are
twisted by the rotating disk to the azimuthal direction,
extracting angular momentum from the disk, enabling
efficient accretion of disk plasmas onto BH and
forming bipolar relativistic jets that are also
collimated by the magnetic force) was proposed not only
for AGN jets (\citealt{Lovelace3,bland3,Pelletier,Meier})
but also for protostellar jets (\citealt{Pudritz,uchida,shibata3,Pudritz2}).
The studies on the dynamical formation of relativistic outflows (jets) (
e.g. \citealt{odell,Begelman,Sikora,Phiney}) from highly luminous
radiation sources, such as AGNs or compact galactic objects show the
connection of jet formation with GRBs in quasars/microquasars
(\citealt{Mirabel,Ferreira23,FerreiraUltra}) since electrons and photons are
coupled due to Thompson Scattering (\citealt{Peebles1,Peebles2,Peebles3,Widrow}).
Recent studies \citep{yuan} indicate that observed AGN jets
are driven by the \citealt{bland1} mechanism.

In (\citealt{yso}) the theoretical model for the disk-jet structure
formation for YSOs was developed  based on the Beltrami Flow model of
(\citealt{SY2011}) using the turbulent viscosity (\citealt{Shakura1973})
as the main reason of accretion; disk was assumed un-magnetized
(no pre-existed global magnetic field was considered).
Analytical conditions for disk-jet structure formation
and parameter ranges for jet-launching and collimation for
YSO Jets were found. The derived solutions describe
the astrophysical disk-jet structures with low ionization, where
the main energy source of the outflow should come from non-magnetic
processes. It was shown that formed disk-jet structure
depends on the thermal properties of disk-flow and
the outflow local Mach number depends on the background
pressure in the jet area.

In the present paper we extend the theoretical study of
(\citealt{yso}) for jet-formation from SAD with electron-ion-photon
gas using the generalized turbulent viscosity approach of
(\citealt{Shakura1973}) in which the dissipation includes both
the photon gas and ion gas contributions being the main source
of accretion. Ignoring the self-gravitation in the disk
we constructed the analytical self-similar solutions for the equilibrium
relativistic disk-jet structure characteristic parameters (velocity field,
generalized vorticity, magnetic field, Alfv\'en Mach number, plasma-beta)
in the field of gravitating central compact object for the force-free
condition justified by observations for our system of study. AD
is weakly magnetized and consists of fully ionized relativistic
electron-ion plasma and photon gas strongly coupled to electron
gas due to Thompson Scattering, often met in different binary systems.
Hence, photon gas behaves as a charged fluid and problem
reduces to the study of relativistic three-fluid system;
the effects of compact object magnetic field 
that may influence the disk formation process, and hence, the equilibrium
disk-jet structure formation (\citealt{Hartmann,Ferreira23}) are ignored.

\section{Physical Model and Equations for Disk-Jet Structure}
\label{Model}

In our model accretion disk consists of fully ionised
electron-ion plasma and photon gas strongly coupled to electron
gas due to Thompson Scattering (see e.g.
\citealt{Peebles2,Widrow,Hartmann,Harrison1973} and references
therein) often met in different binary systems. To describe
corresponding astrophysical disk-jet object
we use relativistic equations for three fluids  [below subscripts
$i$ , $e$ and $\gamma$ are used for ions, electrons and photons,
respectively; electron-ion collisions are ignored]
(\citealt{Weinberg,Baierlein,Peebles2}):

\begin{equation}
\rho_{i}\,\frac{d \mathbf{v}_i}{dt}=
-\nabla p_i + \frac{e n}{c}\left(\mathbf{E}+\mathbf{v}_i\times\mathbf{B}\right)
+ \rho_i\nu_i\nabla^2\mathbf{v}_i + \rho_i\nabla\Phi ,
\label{SS1}
\end{equation}

$$
\rho_{e}\,\frac{d \mathbf{v}_e}{dt}= -\nabla p_e
-\frac{e n}{c}\left(\mathbf{E}+\textbf{v}_e\times\mathbf{B}\right)
$$
\begin{equation}
+\frac{4}{3}\frac{\rho_\gamma}{t_{\gamma e}}\left(\mathbf{v}_\gamma
-\mathbf{v}_e\right)-\rho_e\nabla\Phi ,
\label{SS2}
\end{equation}

$$
\frac{4}{3}\,\rho_{\gamma}\,\frac{d \mathbf{v}_\gamma}{dt}=
-\frac{c^2}{3}\nabla\rho_\gamma
$$
\begin{equation}
-\frac{4}{3}\frac{\rho_\gamma}
{t_{\gamma e}}\left(\mathbf{v}_\gamma-\mathbf{v}_e\right)+
\frac{4}{3}\rho_\gamma\nu_0\nabla^2\mathbf{v}_\gamma-\frac{4}{3}\rho_\gamma\nabla\Phi ,
\label{SS3}
\end{equation}
where $\rho_i , \rho_e $ and $\rho_{\gamma}$ are the densities of ion,
electron and photon fluids and $\nu_i$ and $\nu_0 (=2/9n\sigma_T)$
are the ion and photon fluid's turbulent viscosities, respectively \citep{Chan}:
$t_{\gamma e}=1/(nc\sigma_T)$ is the photon-electron collision time,
$n=n_i=n_e$ is the ion (electron) proper number-density for fully ionized gas; $\sigma_T=
6.65\cdot 10^{-25}cm^2$ is the Thomson cross-section; {$\Phi$ is the} gravitational potential
of the central compact object; it is assumed, that the self-gravity of disk can be ignored.

For our problem of interest we ignore the electron inertia and summing up
equations (\ref{SS2}, \ref{SS3}), ignoring electron fluid stresses compared
to those for photon fluid, we arrive to the so called "charged photon"
fluid equation of motion (see \citealt{Harrison1973,Baierlein} and references therein):
$$
\frac{4}{3}\,\rho_{\gamma}\,\frac{d \mathbf{v}_\gamma}{dt}=
-\nabla p_\gamma
$$
\begin{equation}
-\frac{e n}{c}\left(\mathbf{E}+\textbf{v}_e\times\mathbf{B}\right) +
\frac{4}{3}\,\rho_\gamma\nu_0\nabla^2\mathbf{v}_\gamma-\frac{4}{3}\,\rho_\gamma\nabla\Phi ,
\label{SS4}
\end{equation}
where $p_{\gamma} = (c^2/3)\rho_{\gamma}$ is the photon fluid pressure.

Accreting AD (e.g. YSO disk) -- jet structure, according to observations,
is quite a long-lived object and the steady state solutions could well
describe its behavior. Then, from the equations ({\ref{SS1}}, {\ref{SS4}}),
one can write the equation governing the dynamics of the stationary
compressible magnetised multi-component fluid rotating around
a central massive gravitating object as a whole in the following form:
\[
\rho\left(\mathbf{v}\cdot\nabla\right)\mathbf{v} = - \nabla\mathcal{P}
+ \frac{1}{c}\,\mathbf{j}\times\mathbf{B}
\]
\begin{equation}
+ \frac{4}{3}\,\rho_\gamma \left(\nu_0 + \nu_i
\frac{4n\,m_i}{3\rho_{\gamma}} \right)\nabla^2\mathbf{v}
- \rho\,\nabla\Phi ,
\label{SS5}
\end{equation}
where
\begin{equation}
\mathcal{P}\simeq p_i + p_\gamma ,
\label{SS6}
\end{equation}
is the total thermodynamic pressure,
\begin{equation}
\mathbf{j} = e\,n\left(\mathbf{v}_i - \mathbf{v}_e\right) ,
\label{SS7}
\end{equation}
is the current density and we have introduced for the total
density $\rho $ and common velocity ${\bf v}$ for the combined
system of ion and electron-photon fluid the following
\begin{equation}
\rho = \frac{4}{3}\,\rho_\gamma + \rho_i ,
\label{SS8}
\end{equation}
and
\begin{equation}
\rho \mathbf{v}=\frac{4}{3}\,\rho_\gamma\mathbf{v}_\gamma+\rho_i\,\mathbf{v}_i \ .
\label{SS9}
\end{equation}

We need to add here the Maxwell equations, Equations of State and
Continuity Equations for each fluid. To find the Equation for magnetic field
we recall that we consider the disk (or its regions) such that its
magnetic field (large scale) is weak so that its energy is much smaller
than the fluid energy and, hence, back-reaction of the magnetic field on the
turbulent viscosity (that is the reason for the accretion) is small and we can
ignore the Hall term in the equation (\ref{SS5}) (i.e. we apply below
the force-free condition).

Since the photon fluid, as mentioned above,  is strongly coupled with electron
fluid due to Thomson scattering, as if it is a "charged photon" fluid,
we do not distinguish with photons and electrons and simply have:
\begin{equation}
{\bf v}_{\gamma} \approx \mathbf{v}_e=\frac{3}{4}\frac{1}{\rho_\gamma}
\left(\rho \mathbf{v} -\rho_i \mathbf{v}_i \right) \, , \quad
\mathbf{j}=e\,\frac{\rho}{m_i}\left(\mathbf{v}-\mathbf{v}_{\gamma}\right) \ .
\label{SS10}
\end{equation}
Taking the curl of Eq. (\ref{SS2}) and ignoring the heating/cooling of
electron fluid (not important in equilibrium state, specifically when
the generation of short-scale fields is ignored) due to Biermann effect,
we arrive to:
\[
\nabla \times \mathbf{v}_e \times \mathbf{B}=0
\]
leading to following Beltrami condition for charged photon fluid
(electron/photon-fluid is frozen into the magnetic field):
\begin{equation}
\mathbf{v}_\gamma \parallel \mathbf{B} \ , \qquad \mathbf{v}_\gamma=\mu_{\gamma} \mathbf{B}
\label{SS11}
\end{equation}
where the Beltrami parameter $\mu_{\gamma}$ is a scalar function satisfying the condition:
\begin{equation}
\left(\mathbf{v}_\gamma \cdot \nabla \right)\frac{1}{\mu_{\gamma}}=0
\label{SS12}
\end{equation}
and, hence, the condition for the force-free magnetic field can be written as:
\begin{equation}
{\bf j} \times {\bf B} = 0 \ \  \Longrightarrow \ \ \nabla
\times (\rho {\bf v}) \times {\bf B} = 0 \ .
\label{SS13}
\end{equation}
Note, that the equations (\ref{SS5}) and (\ref{SS13}), together with stationary
Continuity equation $\nabla \cdot (\rho {\bf v})=0$ and Maxwell equation
$\nabla \cdot {\bf B}=0$, are the equations describing the one fluid MHD in
the force-free regime, no two-fluid effects like the generation of short-scale
fields (of both magnetic and velocity nature (\citealt{mmns-1,BSM_deg}))
are taken into account; such effects we will study in our future investigations,
as they are important for the Unified Dynamo/Reverse Dynamo
(\citealt{msms1,msms2,RD_deg,RD_2T}) and heating processes (\citealt{mmns-1}),
as well as the catastrophic acceleration/amplification of velocity/
magnetic fields (\citealt{osym1,osym2,mnsy,BS-flow}) leading to the
modification of dissipation effects (hence, of accretion) but not important
for the formation of minimal (universal) disk-jet structure. Hence, knowledge
of composite fluid momentum ${\bf P} = \rho {\bf v}$ for considered AD is fully
sufficient for the definition of magnetic field (large-scale)
in the minimal model. The assumption for force free magnetic field
for the 3-fluid system of disk regions where the large-scale magnetic
energy is smaller than the flow energy made this possible. The opposite limit
constitutes a different problem, which can work for the areas of inner
disk-edge close to the compact object with strong magnetic fields affecting
the disk formation/disk's local magnetic field (see e.g.
\cite{Ferreira08,Ferrario-3} and references therein).

As shown in \cite{yso,SY2011,Yoshida2012} for disk-jet structure formation
stage magnetic field is not necessary –- having a distinguishable viscosity
in the disk is enough to create the accretion, leading to the feeding of
outflow/jet by disk-flow energy/material; magnetic field affects
significantly the outflow/jet collimation and acceleration.
In ideal MHD numerical simulations, it has been found
that the magneto-rotational instability (MRI) generates turbulence
\citep{BH91}, unless strong non-ideal effects in extremely low-ionized
gases suppress MRI \citep{Bai}. Effect of the magnetic field on the local
turbulence (that amplifies it) may become substantial only when magnetic
energy grows so that it becomes comparable to the kinetic energy of
the small-scale turbulence (important for disk-jet structure formation era) \citep{Baierlein};
then we may expect that in the minimal model of force-free field the formed
disk-jet structure will remain similar to un-magnetized case. One can expect, that
the two-fluid effects (\citealt{mmns-1,msms2}) in disk (arising from \
${\bf j}\times {\bf B}\neq 0$ \ -- when matter is not current-less)
will effectively modify the local accretion due to the creation
of large short-scale magnetic fields feeding turbulence/anomalous
viscosity (\citealt{Baierlein,mmns-1}); also in equilibrium the
Biermann battery effect can be neglected \citep{Widrow} for
electron-fluid (as performed above), as well as the effects of
heating/cooling. These effects as well as the non-force-free regimes
for corresponding disk-regions we will consider in our future studies.

Then, problem of finding the magnetic field for our scenario is reduced
to the problem of finding the fluid-flux/momentum ${\bf B} \sim \rho {\bf v}$ --
the large-scale magnetic field behaves similar to the total current (flux for
the flow, that is a momentum \ ${\bf P}$). We need to find the behaviour of velocity field
and formulate the constraint on Beltrami parameter for entire composite flow:
$\rho {\bf v} = \mu {\bf B}$ \ with \ $\mu \gg 1$ ; the results of opposite
range ($\mu \ll 1$ for the strongly magnetized disk case) will be studied elsewhere;
working out the consequences of related phenomenon for disk-jet structure
formation is underway.

It is interesting to note, that the governing equations for our
geometrically thin disk with rotating composite system of ion fluid
and "charged photon fluid" in stationary state are reduced to the equations
that describe the neutral rotating disk matter around the compact
object studied in \cite{yso}, with \ $\rho, \ {\bf P}, \ \nu , \ {\mathcal P}$ \
representing the characteristic parameters of composite fluid and the
accretion caused mainly due to the photon fluid viscosity
(although the generalized viscosity includes both the photon
gas contributions as well as of ions the role of ions relative to photons
is negligible in the local turbulent-viscosity (see e.g. \citealt{Harrison1973,Chan,Baierlein}).
Then, we can follow the above mentioned paper to find the velocity
field of a disk fluid (as a whole) solving following equations:
\begin{equation}
({\bf v} \cdot \nabla) {\bf v} = - \nabla h - \nabla \Phi +
{1 \over \rho} \nabla \cdot {\rm T} \ ,
\label{PDE1}
\end{equation}
\begin{equation}
\nabla \cdot (\rho {\bf v} ) = 0  \ ,
\label{PDE2}
\end{equation}
where ${\bf v}$, $\rho$ and $h$ are the velocity, density and enthalpy
of composite system, respectively; we have used the barotropic
equation of state (for our problem of study) to calculate the
enthalpy of the fluid:
\begin{equation}
\nabla h = {1 \over \rho} \nabla {\cal{P}} ~,
\label{PDE3}
\end{equation}
where the viscous stress tensor \ { $T_{ik}$ (now representing the
total dissipative effects defining the accretion) and the corresponding
term in Eq.(\ref{PDE1}) is formally written as}:
\begin{equation}
\nabla \cdot {\rm T} \equiv \nabla_k \, T_{ik} = {\partial \over
\partial x_k} T_{ik} ~.
\label{SS14}
\end{equation}
Following \cite{SY2011,Yoshida2012}, introducing a so-called ``ideal''
and ``reduced'' factors of ``local'' density (see also Arshilava et al 2019)
\begin{equation}
\rho = \rho_{I} \rho_{R} ~
\label{SS15}
\end{equation}
to separate the ideal fluid and the dissipative effects and to track
the accretion effects in disk-jet structure formation process so that
$\rho_{R} = 1$ and $\rho_{I} = \rho$ in conventional ideal fluid mechanics
with zero dissipation, we seek the steady state solutions of the disk-jet
structures persisting around a central accreting slowly rotating
compact object: The details of derivation
of major analytical relations are presented in Appendix A.

Then, the stationary state of the rotating thin composite system of AD
can be fully investigated using Eqs. (\ref{SS13},\ref{Beltrami1},\ref{Bernoulli21},
\ref{Bernoulli22}) and the explicit form of the viscous stress tensor related
to the specific object conditions. In present paper we employ the model in which
the small scale turbulence creates the anomalous dissipation that can be
described by using the generalized $\alpha$--viscosity model introduced by \cite{Shakura1973}
in which we use the effective viscosity model both in the disk and
jet as well as in the disk-jet transition areas (\citealt{yso}).
Assuming the
strong azimuthal rotation, splitting the pressure $\cal{P} $ into the
background constant component ${\cal P}_0$ and deviation form it $p$ ,
\begin{equation}
{\cal{P}} = {\cal P}_0 + p \ ,
\end{equation}
the turbulent stress tensor can be split into the background
constant component $\bar T_{ik}$ and smaller varying deviation $t_{ik}$:
\begin{equation}
T_{ik} = \bar T_{ik} + t_{ik}
\label{Tik}
\end{equation}
with $\bar T_{r \varphi} = \alpha_0 {\cal P}_0 $,
assuming axisymmetric flow $v_\varphi = r \Omega_{\rm K}(r,z)$ rotating
locally with Keplerian angular velocity (justified for the rotationally
supported flow, for which the radial pressure gradients can be negligible compared to
the centrifugal force -- situation found for slowly accreting flows
with slowly varying background pressure ${\cal P}_0 \simeq const $ ):
\begin{equation}
\Omega_{\rm K}^2(r,z) = {G M_\star \over (r^2 + z^2)^{3/2}} ~, \label{Omega_Kep}
\end{equation}
where $M_\star$ is the mass of the central object, in the axisymmetric
case, the only significant viscous stress tensor
elements can be calculated as follows:
\begin{equation}
t_{r \varphi} = {r^2 \over r^2 + z^2} \,\beta \,p ~, \label{Tr}
\end{equation}
\begin{equation}
t_{z \varphi} = {r z \over r^2 + z^2}\, \beta  \,p ~, \label{Tz}
\end{equation}
where positive/negative $p$ corresponds to the stronger/weaker
turbulence compared to the background generalized turbulent steady state.

The geometry of the observed disk-jet structures and the continuity
Eq. (\ref{PDE2}) dictate the expansion of the flow velocity
in the following way (\citealt{SY2011,yso}):
\begin{equation}
{\bf v} = {1 \over \rho} \left( \nabla \psi \times \nabla \varphi \right)
+ r v_\varphi \nabla \varphi ~,
\label{Stream}
\end{equation}
where the axisummetric stream function $\psi$ exactly matches the stream
function of the actual momentum ${\bf P} = \rho {\bf v}$ of the bulk flow.
Then, the straightforward algebra presented in Appendix A leads to
the Generalized Beronoulli Condition and Beltrami Condition
represented by the Eqs. (\ref{Stream1},\ref{Stream2},\ref{E}) and (\ref{beltrami2})
which constitute the final set of equations.

\section{The similarity solutions for disk-jet structures}
\label{Similarity}

To construct the solutions representing disk-jet structure,
following \cite{SY2011}, we introduce the orthogonal variables
$\tau = z/r$ and $\sigma = \sqrt{r^2 + z^2}$ so that
$\nabla \tau \cdot \nabla \sigma = 0$ leading to the obvious
results for the Gravitational Potential to be $\Phi=\Phi(\sigma)
= - \, \Omega_0^2/\sigma $ and the stream function to be dependent
only on the $\tau$ variable \ ($\psi = \psi(\tau) $):
thus, making it possible to separate the variables in the solution.

Then, after some straightforward algebra (see Appelndix A), when
seeking solution of the system assuming that describing physical variables
($p, \rho_{\rm I}, \rho_{\rm R}, \lambda $) can be factorized
using the similarity variables $(\sigma, \tau $)
applying variable splitting ansatz (like e.g. $p(\sigma,\tau) =
p_1(\sigma) p_2(\tau) $, see (\citealt{SY2011,yso})),
for the azimuthal velocity we obtain:
\begin{equation}
v_\varphi(\sigma,\tau) = V_{\rm Kep} = {\sigma_0 \Omega_0 \over (1+\tau^2)^{1/2}}
\left( {\sigma \over {\sigma_0}}\right)^{-1/2} ~, \label{Kepler}
\end{equation}
where $\Omega_0$ is the Keplerian angular velocity of the rotation at some
characteristic radius $\sigma_0$ in the disk.
Then, applying this ansatz we find
the explicit solutions for all ($p, \rho_{\rm I}, \rho_{\rm R},
\lambda $) \ in the $\sigma$ coordinate. After finding the radial profiles
of these parameters
one derives the defining algebraic equation (\ref{Realizability}) which
links the values of
$\beta$ parameter (defining the generalized viscosity of our composite
3-fluid system), $\tau$-dependent part of density $\rho_2$ and the stream-function $\psi$,
and represents the "realizability" condition for all existing solutions
within the considered Beltrami flow model of disk-jet structure formation
(the details can be found in Appendix A).

\begin{figure}
\begin{center}
\includegraphics[width=0.8 \columnwidth]{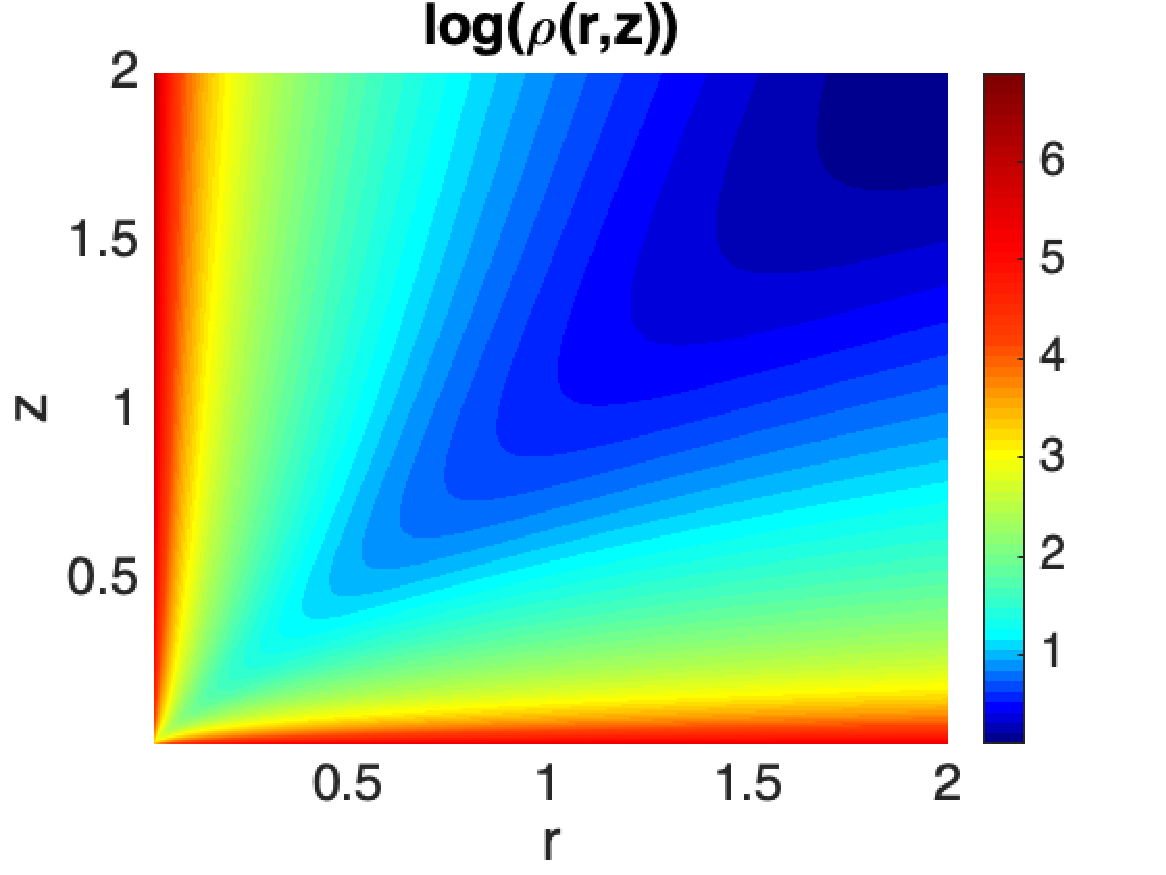}
\end{center}
\caption{Total density $\rho(r,z)$ distribution of the 
weakly magnetized disk-jet structure for:
$A_d=3$, $A_j=3$, $m_d=-3$, $m_j=3$ and $\tau_0$ = 0.01;
for $\rho_2(\tau)$ the power-law distribution (\ref{Rho2SY}) was used
(compare with density distribution [Figure 4] of the unmagnetized
disk-jet structure of \citep{yso}.}
\label{surfs}
\end{figure}

Then, applying the same methodology as studied in \citep{yso}
the general solutions of our weakly magnetized
disk-jet model can be easily calculated - (see the
Eqs. (\ref{Solrho}-\ref{Sol2})) which together with the radial profiles
and appropriate choice of the solution for $W$ give the full
solution for the disk-jet bulk (ion and electron-photon)
flow for different types of density profiles $\rho_2(\tau)$ for
which we employ the power-law distribution (cf. \citealt{SY2011}).
Figure \ref{surfs} shows the total density (dimensionless)
distribution of the weakly magnetized disk-jet structure.
One can imagine that constructing the solution for only ion
density one would find it different (in numbers) from the total density,
although, geometrically similar to the total density presented above
since the considered model dictates such solution.

Finally we obtain the velocity field components of the
weakly magnetized disk flow (of composite
3 -fluid system consisting of ion, electron and photon gases):
\begin{equation}
v_{rD}(\sigma,\tau)  = -{2 \over 5} {\tau^2 \over (1 + \tau^2)^{3/2}}
\left( {\sigma \over \sigma_0} \right)^{-1/2} \beta \sigma_0 \Omega_0~,
\label{Sol3}
\end{equation}
\begin{equation}
v_{zD}(\sigma,\tau)  = -{2 \over 5} {\tau^3 \over (1 + \tau^2)^{3/2}}
\left({\sigma \over \sigma_0}\right)^{-1/2} \beta \sigma_0 \Omega_0 ~,
\label{Sol4}
\end{equation}
and the jet flow:
\begin{equation}
v_{rJ}(\sigma,\tau) =  {5 \over 2} {1 \over (1 + \tau^2)^{1/2}}
\left( {\sigma \over \sigma_0} \right)^{-1/2} \frac{\sigma_0 \Omega_0}{\beta} ~,
\label{VrJ}
\end{equation}
\begin{equation}
v_{zJ}(\sigma,\tau)  =  {5 \over 2} {\tau \over (1 + \tau^2)^{1/2}}
\left( {\sigma \over \sigma_0} \right)^{-1/2}  \frac{\sigma_0 \Omega_0}{\beta} ~.
\label{VzJ}
\end{equation}

\begin{figure}
\begin{center}\includegraphics[width= 0.98 \linewidth]{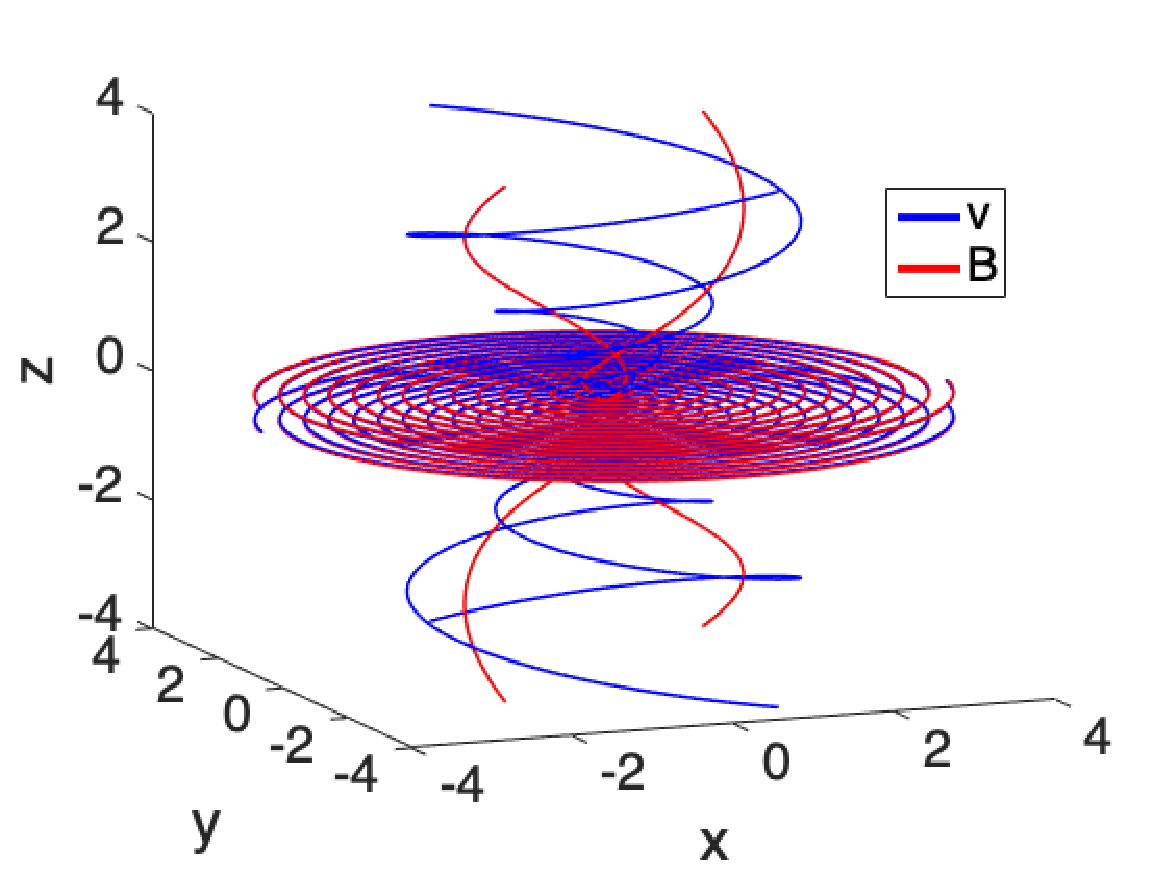}\end{center}
\caption{Velocity streamlines (blue) of the disk-jet structure illustrating
accretion-ejection flow and global disk-jet magnetic field (red)
at $\tau_d=1$, $\tau_j=2$, $\tau_0=0.01$ $\beta=0.01$, $\kappa = 0.001$.
Swirling of the velocity field is stronger than that of the magnetic field
for the weakly-magnetized disk assumption.}
\label{velo+mag}
\end{figure}

Estimating the accretion speed of the flow in the disk region ($\tau
< \tau_d$):
\begin{equation}
v_{\rm acc} = \left(v_{rD}^2 + v_{zD}^2 \right)^{1/2} =
{2 \over 5} \ \beta V_{\rm Kep} ~ {\tau^2 \over (1+\tau^2)^{1/2}}  ~,
\label{Vacc}
\end{equation}
and ejection velocity in the jet region ($\tau > \tau_j$):
\begin{equation}
v_{\rm ej} = \left(v_{rJ}^2 + v_{zJ}^2 \right)^{1/2} =
{5 \over 2} {V_{\rm Kep} \over \beta}  ~
\left( 1+\tau^2 \right)^{1/2} ~.
\label{Vej}
\end{equation}
we conclude, that in the low $\beta$ limit derived solution
corresponds to the locally slowly accreting flow
($v_{\rm acc} \ll V_{\rm Kep}$) in the disk with the
locally fast outflow in the jet ($v_{\rm ej} \gg V_{\rm Kep}$),
matching the properties of astrophysical accretion-ejection flows.
Notice, that above expressions do not depend on the explicit profile
of $\tau $-dependent part of density (see \citealt{yso}). Then, the
continuous velocity field of bulk flow (as well as the continuous
magnetic field via the relation (\ref{SS13})) can be calculated using
Eqs. (\ref{SolVr}), (\ref{Sol2}) with the three region solution for the
$W(\tau)$ function. For the magnetic field solutions
we can employ stationary Continuity Equation \ $(\nabla \cdot \rho {\bf v})
= 0$ \ and \ $\nabla \cdot {\bf B} = 0$, which, together with \
$\rho {\bf v} = \mu {\bf B}$ \ with \ $\mu \gg 1$ \ for our weakly
magnetized disk yield
\[
({\bf v} \cdot \nabla)\frac {1}{\mu} = 0 \qquad \Longrightarrow
\qquad \frac{\partial}{\partial \tau} \mu = const
\]
leading to \ $\mu^{-1} \simeq \kappa \,\tau $ \ with \ $\kappa \ll 1$ \ for
our model (with variables splitting ansatz for similiraty solutions).
Figure \ref{velo+mag} illustrates the streamlines of composite flow (blue)
and global magnetic field (red) of the derived weakly magnetized disk-jet structure.
Notice, that the construction of Velocity field doesn't require knowledge of density
profile while magnetic field in our weak-field approximation with force-free
condition needs the density-profile to be known; also, swirling of the composite
flow is faster than that of the magnetic field -- this could be expected since
only electrons are coupled with photons and not the entire e-i plasma,
difference in the corresponding characteristic length-scales for ions and electrons
shall cause this departure. It is interesting that numerous numerical
runs showed no dependence of swirling ratio on $\kappa $ for these vector-fields.

\begin{figure}
\begin{center}
\includegraphics[width=0.90 \columnwidth]{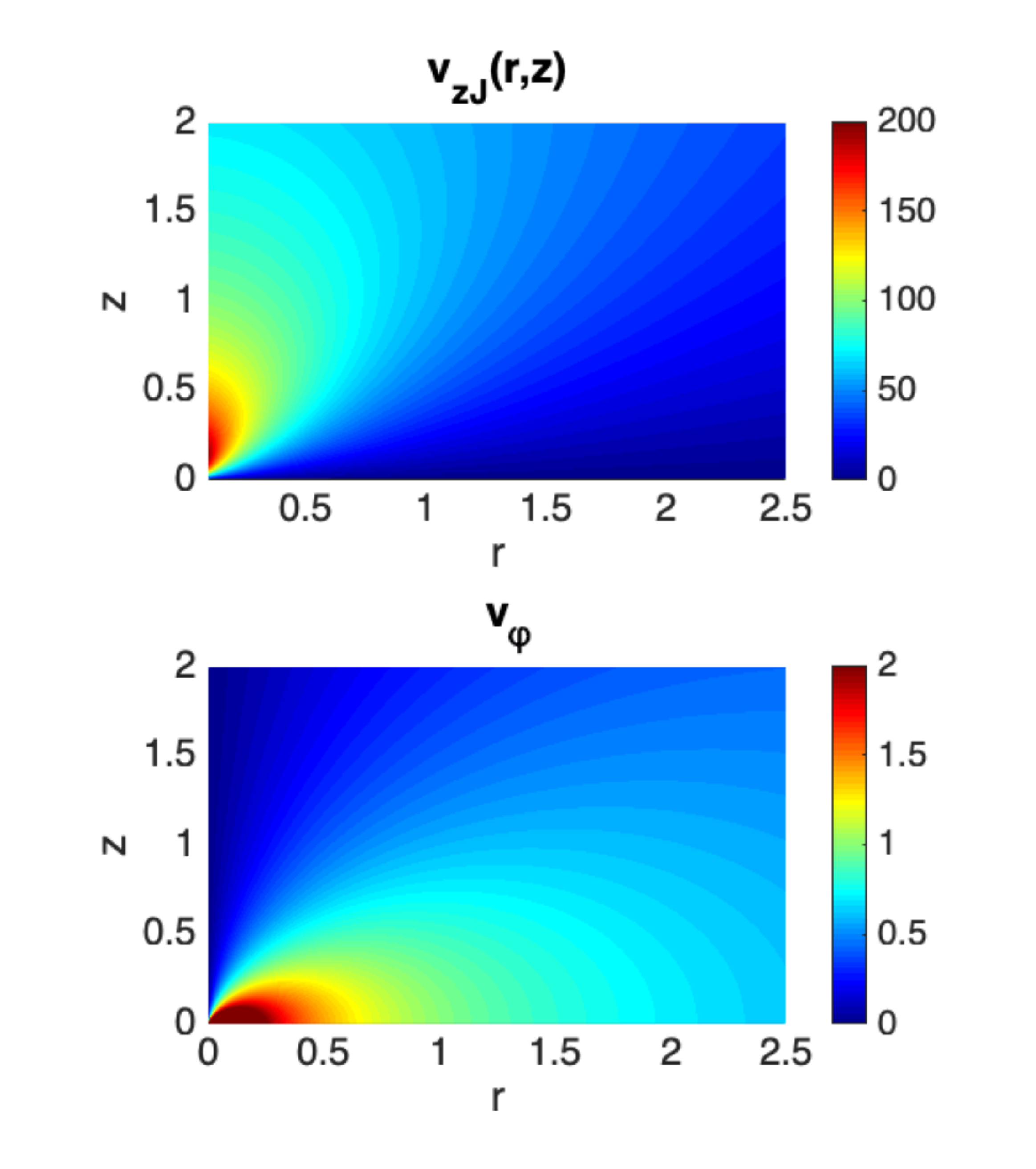}
\end{center}
\caption{Profiles of the vertical velocity of the relativistic jet solution
$v_{zJ}(r,z) / ( 2.5 \sigma_0 \Omega_0)$ (see Eq. (\ref{VzJ})) (top)
and the azimuthal velocity $v_{\varphi}=V_{Kep} / (\sigma_0 \Omega_0)$
in the disk (bottom) for $\beta = 0.01$.
Maximal velocity of the outflow is reached near the vertical axis above
the disk plane, at the top edge of TR while the maximal value of azimuthal
(Keplerian) velocity is reached near the inner edge of a disk at the
bottom edge of the TR at the disk mid-plane.
\label{vz_vphi}}
\end{figure}

Disk-Jet structure configuration we constructed in present paper is
a large-scale equilibrium structure the formation of which, as shown
in \citep{yso}, is guaranteed for accretion-ejection flows with turbulent
viscosity; the inclusion of weakly magnetized disk and photon gas effects
extends the model for wider class of disk-jet structure systems.
Observations also show, that geometrically all disk-jet structures show
the universal geometrical character; the collimation efficiency of the jet
is significantly defined by the acceleration mechanisms and magnetic field
-- we emphasize here, that the formation of jet and its further acceleration
are different stages in the dynamics of the jet-evolution and, as shown
in \citep{SY2011,yso}, for the minimal model there is no need in magnetic
field for the jet-formation era.

\section{Characteristic physical parameters of the constructed
weakly magnetized disk--relativistic jet structure}
\label{Properties}

In this section we present the illustrations for the characteristic
parameters of the weakly magnetized disk--relativistic jet structure
based on the solutions derived in the present paper emphasizing that the
specific value of the local generalized viscosity $\beta $ parameter can be
found from observations for the concrete disk-jet object for which
both the radial accretion and the vertical ejection velocities
nearby the central object could be measured. From the Eqs.
(\ref{Vacc}) and (\ref{Vej}) we find $\beta^2 \sim {v_{\rm acc} /
\,v_{\rm ej}}  $ and since the value of $\beta \, (< \alpha_0)$
parameter is constrained by $\alpha_0$ \, (the parameter
describing anomalous viscosity due to background stationary
turbulence) one easily finds:
\[
v_{\rm ej} > 10^{4} \ v_{\rm acc}
\]
using a typical value from observational luminosity $\alpha_0 \sim 0.01$
(see \cite{yso} and references therein).

\begin{figure}
\begin{center}
\includegraphics[width=0.9 \columnwidth]{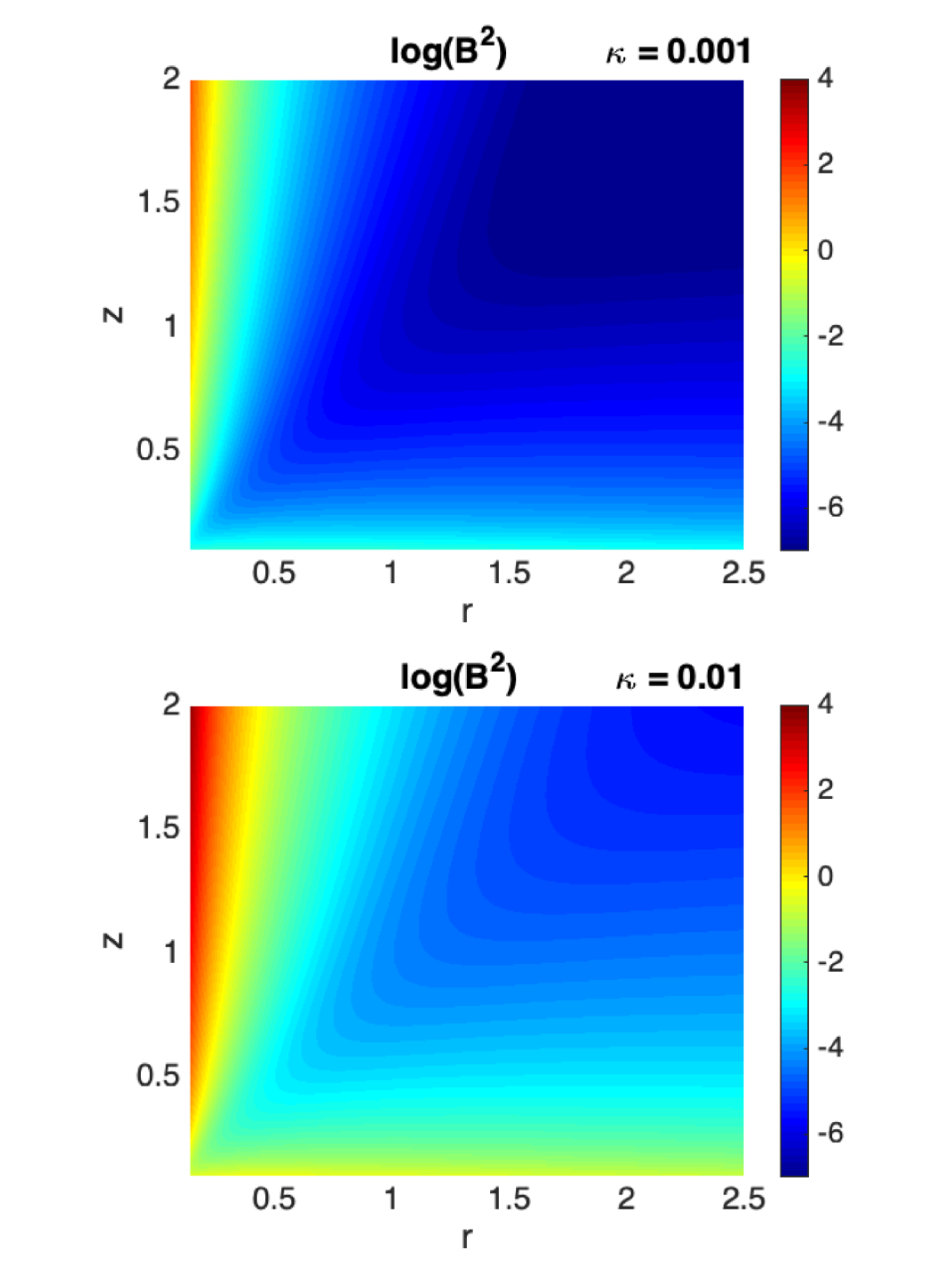}
\end{center}
\caption{Magnetic Energy [${\rm log}({\bf B}^2)$] normalized
to $p_0$ for \ $\beta = 0.01$ \ plotted for different \ $\kappa $-s.
The accumulation of magnetic energy in the Jet region is
clearly seen reaching maximal values at the jet (vertical) axis;
process is stronger for larger $\kappa $; the jet region
magnetic field is several orders greater compared to its disk area values.
For density the distribution of Figure \ref{surfs} was used.}
\label{MagnEn}
\end{figure}

To illustrate the properties of our solutions for the weakly magnetized
disk-relativistic narrow jet structure the vertical velocity distribution
of the jet flow and the azimuthal velocity are plotted in Figure
\ref{vz_vphi} (see Eqs. (\ref{VzJ}) and (\ref{Kepler})) for
$\beta = 0.01$ -- compare with Figure 5 of \citep{yso} (for which $\beta = 0.002$
was used and not the indicated in its Figure caption parameter). The outflow/jet
launching is at the bottom edge of transition region (TR), just above
the disk-surface and reaches its maximal value at the top edge of TR
beyond of which the vertical flow velocity $v_z$ decreases both with
vertical and radial distances, similar to the Keplerian profile
($\propto (\sigma / \sigma_0)^{-1/2}$). Hence, the solutions derived
within the minimal Beltrami-Bernoulli flow model well describes
the {\it formation} of the weakly magnetized disk-relativistic jet
structure (similar to un-magnetized case of (\citealt{SY2011,yso})).
We remind the reader that in our model the effects of central object
magnetic field/jets, disk-winds as well as the heating/cooling
processes were not considered. We believe that invoking these effects as well
as the two-fluid effects (as discussed above) the generalized Magneto-Beltrami-Bernoulli
mechanism (\citealt{mnsy,msms2}) may further accelerate and collimate the jet-flow.
Also, depending on the plasma condition near the central object
the disk-jet connection point may differ (one needs to add
the equation of state taking into account the collisional effects
together with pair creation, erruptive/explosive effects and etc.
(see e.g. \citealt{Widrow,Mirabel,Ferreira23,bland,bland4,zanni,mignone,Mouchet} and
references therein); investigations of these issues are underway.

Constructing the local Mach Number the possibility of forming the
supersonic outflow in the narrow jet region at $\tau \gg 1$ near the
jet axis was discussed in \citealt{yso} where the feasibility of
derived solution with the Laval nozzle mechanism was also shown; similar
result will persist also for our composite electron-ion-photon gas outflow.
Moreover, since the pressure variation is negative in the jet region
($p<0$ for $W_-(\tau)<0$) the swirling solution in the jet region leads
to the decrease of sound speed increasing the local Mach number of
the outflow; when adding the cooling effect (not considered in minimal
Beltrami-Bernoulli flow model) such increase will be stronger widening
the area of supersonic flow existence.

For our weakly magnetized AD model of composite 3-fluid system it is
essential to follow the distribution of the Alfv\'en Mach number
($ = v_z {\sqrt{\rho}}\,/ B = v_z/(\kappa \tau {\sqrt{\rho}}\,v)$ )
and Magnetic Energy throughout the formed disk-jet structure. For this
purpose we constructed the illustrating plots for these 2 physical
parameters using our self-similar solutions; for the background
pressure we used the normalization of the background pressure
${\cal P}_0$ on the pressure $p$ of the self-similar solution
(\citealt{yso}) at $\sigma=\sigma_0$ and $\tau = \tau_j$
(see Eqs. (\ref{Sol1},
\ref{Rho2SY})):
\begin{equation}
p_0 \approx {5 \over 2} {\sigma_0^{1/2} \Omega_0^2 \over
\beta^2} A_j \tau_j^{m_j} ~,
\end{equation}
assuming that $\tau_0 \ll \tau_j$.

\begin{figure}
\begin{center}
\includegraphics[width=0.9 \columnwidth]{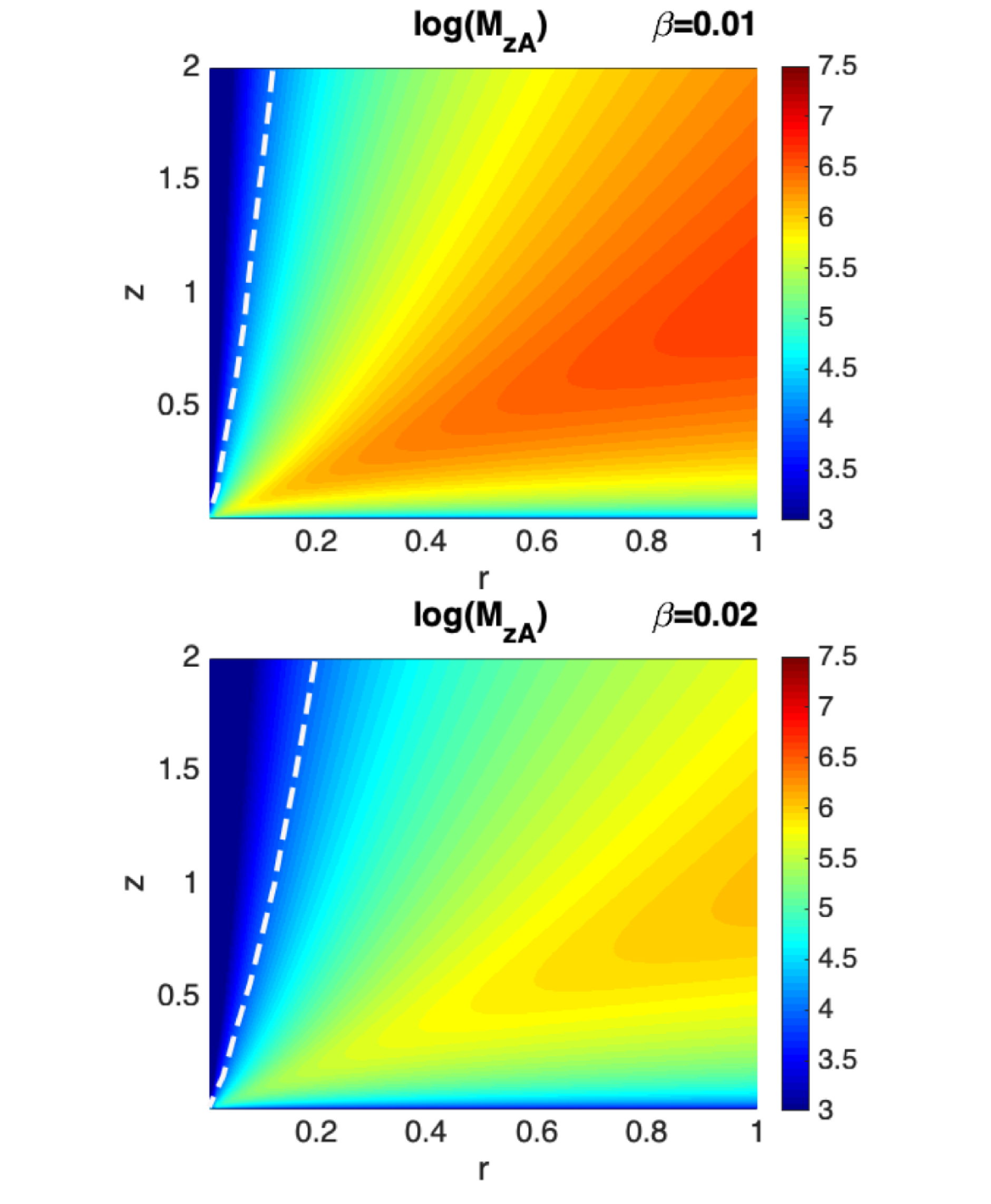}
\end{center}
\caption{Vertical Alfv\'en Mach number of the jet flow \
[${\rm log}(M_{zA}(r,z))$] \ when \ $\tau_0=0.01$, $\kappa = 0.01$,
$\sigma_0=1$ \ for different \ $\beta$-s ($=  0.01, \, 0.02$
\ from top to bottom). For density the distribution presented in
Figure \ref{surfs} was used. The flow is super-Alfv\'enic
\ ($M_{zA} \gg 1$) \ everywhere decreasing with \ $\tau \gg 1$; \ with larger
\ $\beta $ \ the area of jet with \ $M_{zA}(r,z) < 10^4$ \
(see dashed line) gets wider; \ $M_{zA}(r,z)$ \ decreases along vertical
axis indicating that radiation exists beyond the outflow
major concentration span (compare with Figure \ref{vz_vphi} ).}
\label{MzA}
\end{figure}

\begin{figure}
\begin{center}
\includegraphics[width=0.95 \columnwidth]{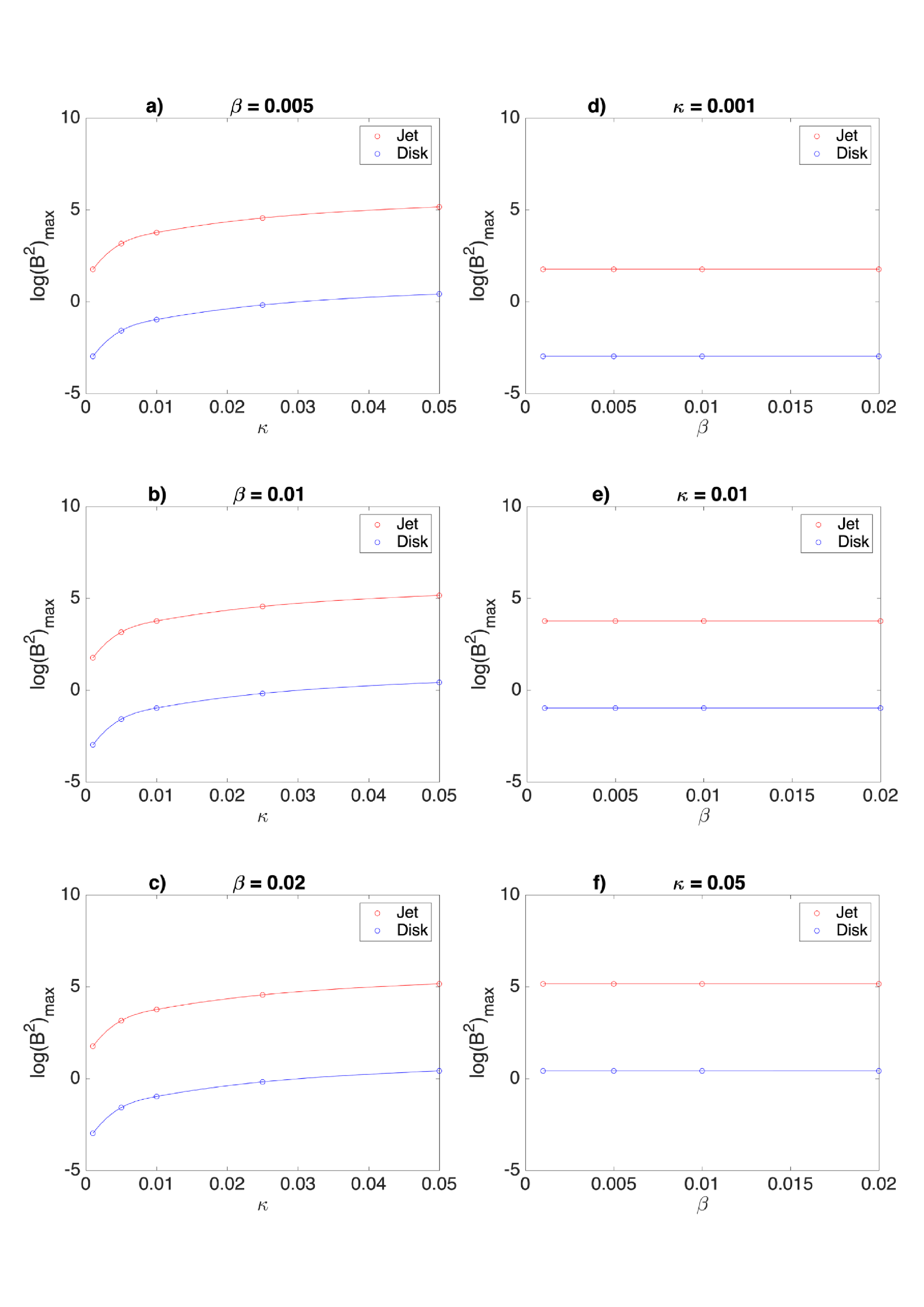}
\end{center}
\caption{Maximal values of magnetic energy in jet (red) and disk (blue)
for different turbulent generalized viscosity coefficients \
($\beta = 0.005, \, 0.01, \, 0.02$) versus \ $\kappa $ \ (see plots
a), \, b) \, c) of the left column) and for different \ $\kappa $-s \
versus $\beta $ (see plots d) \, e) \, f) of the right column);
magnetic energy is several orders larger in jet compared to its
value in the disk for all cases. Picture is not sensitive to changes
in \ $\beta $ \ for the same \ $\kappa $ \ while for the same \ $\beta $ \
it increases gradually with $\kappa $. }
\label{maxvalues}
\end{figure}

Figure \ref{MagnEn} shows the distribution of magnetic energy throughout
the equilibrium Disk-Jet structure. One observes the accumulation of
magnetic energy in the narrow region of formed Jet (near its axis),
this process is stronger for bigger $\kappa $ reflecting the effect
of initial conditions in the electron-ion-photon gas accretion disk;
also note, that the jet region magnetic field is several orders
greater compared to its disk area values.
The plots for vertical Alfv\'en Mach number are illustrated in
Figure \ref{MzA} for different turbulent generalized viscosity parameter
$\beta $ for the specific $\kappa = 0.01$. One observes that the flow
is super-Alfv\'enic ($M_{zA} \gg 1$) everywhere while its strength
compared to magnetic energy reduces at $\tau \gg 1$;
with larger $\beta $ the area of jet with $M_{zA}(r,z) < 10^4$
(see dashed line) gets wider; $M_{zA}(r,z)$ decreases along
vertical axis indicating that radiation exists beyond the
outflow major concentration span. In Figure \ref{maxvalues} the
maximal values for magnetic energy in the jet (red) reached at
the vertical axis and disk (blue) reached at the mid-plane of the
disk are illustrated: (i) versus $\kappa $ for different
turbulent viscosity parameter $\beta $ ({\it left panel})
and (ii) versus $\beta $ for different $\kappa $-s ({\it right panel}).
One clearly observes that the maximal values of
magnetic energy are several orders larger in the jet compared
to its values in the disk for all chosen cases. Moreover,
the result is not sensitive to changes in $\beta $ for
the same $\kappa $ while for the same $\beta $ it increases gradually
with $\kappa $ -- clear link to the initial preparation of the
electron-ion-photon gas accretion disk / concrete conditions of
specific astrophysical object.

Since in our model the effect of the central object magnetic field
was not considered, it is expected that the final results for
observed Alfv\'en Mach number (as well as the Magnetic Energy)
in the jet region may decrease (increase) in the realistic
electron-ion-photon disk-relativistic jet systems; also the
decrease of the generalized $\beta $ parameter leads to the
increase of the ratio between the vertical and radial velocities
and, consequently, change of the equilibrium disk-jet
flow/magnetic field geometry for various astrophysical objects.
Even then, interestingly, the solutions derived in the minimal
model considered in the present work applying Generalized
Beltrami-Bernoulli flow model for equilibrium disk-jet structure,
can well mimic slow radial sub-Keplerian accretion composite flow
in the weakly magnetized disk region and fast narrow super-Keplerian
collimated outflow with strong magnetic field in the jet region.
As already discussed, inclusion of the above effects together with
the Hall effect for stronger magnetic field regions will make
the constructed disk-jet structure characteristic parameters
closer to the realistic/observed astrophysical objects; prediction
of the jet parameters will be more reliable as well.

\begin{figure}
\begin{center}
\includegraphics[width=1.00 \columnwidth]{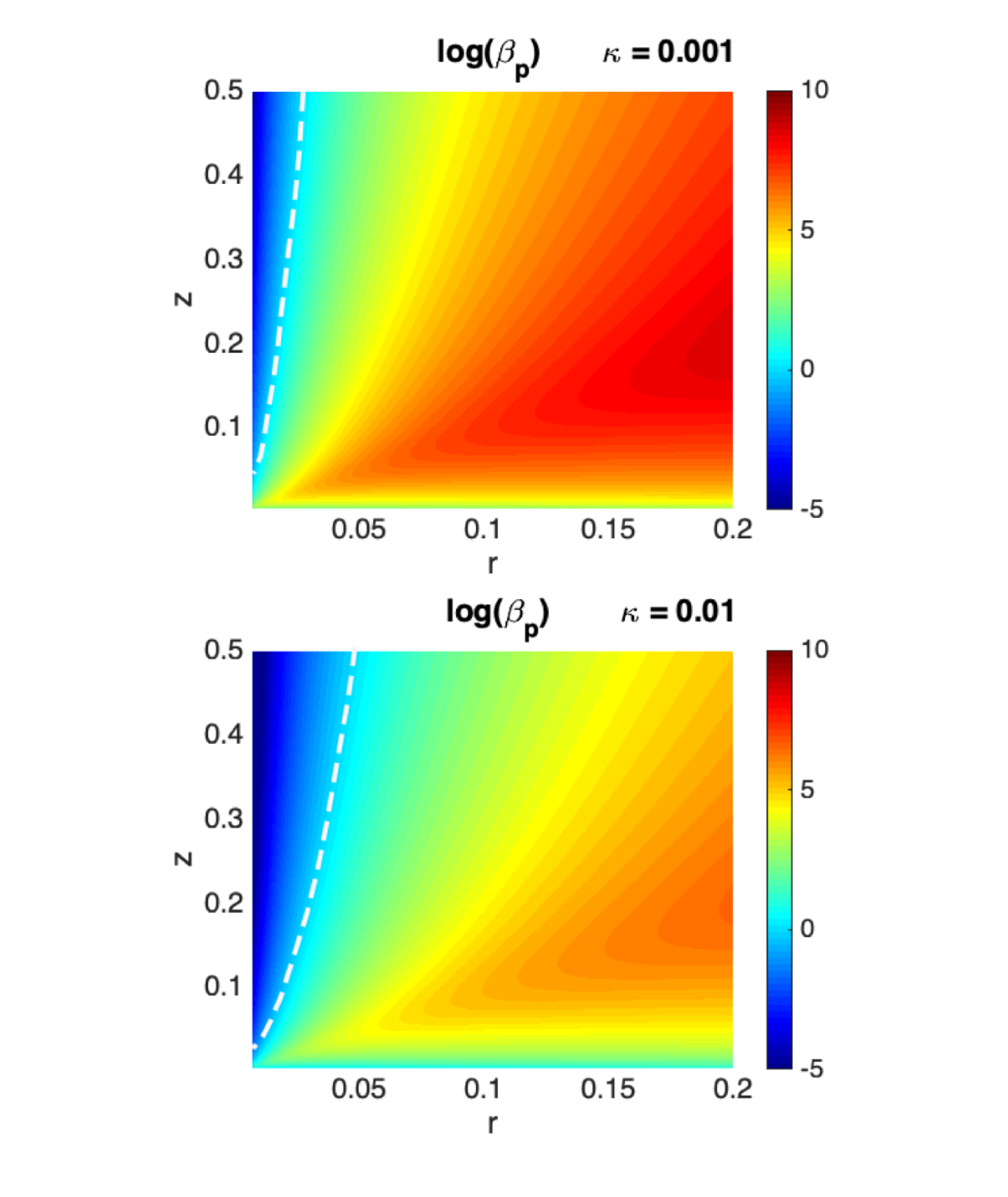}
\end{center}
\caption{Plasma-beta \ $\beta_p$ \ for turbulent generalized viscosity
parameter \ $\beta = 0.01$ \ for \ $\kappa = 0.001$ (top), \, and \
$\kappa =0.01$ \ (bottom) when \ $\tau_0=0.01$, \ $\beta = 0.01$; \
$\sigma_0=1$ \ and \ ${\cal P}_0/p_0 = 10^{5}$ ; \ the dashed line
represents the place of \ $\beta_p =1$ \ beyond which towards jet-axis \
($\tau \gg 1$) \ the outflow \ $\beta_p < 1$ .
For density the distribution presented in Figure \ref{surfs} was used.}
\label{PlasmaBeta}
\end{figure}

To better illustrate the reliability of our approach we give the results
for local plasma-beta parameter $\beta_p$ in Figure \ref{PlasmaBeta}
for the pressure ratio ${\cal P}_0/p_0=10^{5}$ for different local
magnetic field strengths (different $\kappa $-s) justified by
observations. The dashed line represents the place of plasma
$\beta_p =1$ beyond which, towards jet-axis ($\tau \gg 1$),
the outflow plasma-beta $\beta_p < 1$ (being at the same time
supersonic close to axis); at wider angles ($\tau > 1$)
the outflow plasma $\beta_p > 1$ (being subsonic) and, thus,
should be decreasing away from the central object (from
Figure \ref{vz_vphi} we observe \ $v_z(\sigma) \propto \sigma^{-1/2}$)
(compare e.g.  with \cite{Acre2013,Lee2018}).
Notice, that bigger the local magnetic field strength ($\kappa $) larger
is the area with plasma $\beta_p < 1$ in the jet (see e.g.
\citealt{Begelman,Vlemmings,Spruit1,Spruit2,Matsumoto1,Matsumoto2,Ferreira23,Koide}
and references therein) for fixed viscosity parameter $\beta $. In fact,
this effect shall become significant with inclusion of cooling
(effect not considered in present approach) that will further
reduce the local sound speed leading to the decrease in the
local plasma-beta (with simultaneous increase of
local Mach number (\citealt{yso}). This effect could add to the
jet acceleration due to the farther adiabatic expansion; while
in opposite limit the increase of the background pressure ${{\cal P}_{0}}$ would
increase local plasma-beta (decrease the local Mach number to the subsonic
values) changing the final geometry of disk-jet structure.

\section{Summary}
\label{Summary}

We extend the theoretical Beltrami-Bernoulli flow model
of (\citealt{yso}) for the relativistic jet-formation from SAD
with electron-ion-photon gas using the Generalized Turbulent
Viscosity approach (\citealt{Shakura1973}) in which the dissipation
includes both the photon gas and ion gas contributions being
the main source of accretion in the weakly magnetized disk
while the star-formation process. Ignoring the self-gravitation
in the disk we derived analytically the magnetized collimated
jet outflow self-similar solutions for the force-free condition:
for the stationary turbulent state (powering the accretion process
both in the disk and jet of the unified equilibrium disk-jet structure
in the field of gravitating central compact object) we constructed the
characteristic parameters like velocity field, generalized vorticity,
magnetic field, Alfv\'en Mach number, plasma-beta.

In our model the accretion disk is weakly magnetized and
consists of fully ionized relativistic electron-ion plasma and
photon gas strongly coupled to electron gas due to Thompson
Scattering, often met in different  binary systems. Hence,
photon gas behaves as a charged fluid and problem reduces
to the study of relativistic three-fluid system; the effects
of compact object magnetic field that may influence
the disk formation process, and hence, the equilibrium
disk-jet structure final geometry, Hall term, as well as
the heating/cooling and ionization modifying the local
accretion were ignored. Also, in present analysis magnetized
disk-winds were not invoked; the additional dissipation
mechanisms as well as the phenomena in the vicinity of
central object may play significant role too for the
final physical parameters of formed disk-jet structure.
We believe, that dynamical effects together with all above
physical phenomena will make the solution of the problem
only richer; the consequent problems of jet further
acceleration as well as heating could be considered too
(see e.g. (\citealt{osym1,osym2,mnsy,msms2}) for the possible
jet-flow acceleration due to Hall effect in the generalized
magneto-Bernoulli mechanism); study of these issues
are planned for our future investigations.

Similar to un-magnetized case (\cite{yso}) our relativistic
magnetized disk-jet structure parameters depend on the
thermal properties of the disk flow -- the local Mach number
of the jet-outflow depends on the background pressure in the
jet area and at low pressure, jet outflow reaches supersonic
amplitudes close to the central axis of the flow being transonic
from low poloidal angles to higher poloidal angles decelerating
with radial distance. Interestingly, the local Plasma-beta $\beta_p $
shows opposite distribution -- being $> 1$ in the disk it passes
the value $1$ at low poloidal angles and closer to jet axis it
becomes $< 1$; such area is wider for bigger local background
disk-magnetization ($\kappa $) (see e.g.
\citealt{Vlemmings,Spruit2,Matsumoto2,zanni,Koide}) for fixed local
viscosity. We observed the accumulation of magnetic energy in the
narrow region of formed Jet (near its axis), being several orders
greater compared to its disk area values; the result is not sensitive
to changes in local generalized turbulent viscosity parameter $\beta $
for the same background magnetic fields while for the same $\beta $
it depends on the background field in disk -- this process is stronger for
higher background magnetization of the disk reflecting the dependence
on the initial conditions in the electron-ion-photon gas accretion disk.
The results for vertical Alfv\'en Mach number show that the flow
is super-Alfv\'enic ($M_{zA} \gg 1$) everywhere in the equilibrium
relativistic disk-jet structure while the outflow kinetic energy
decreases faster with vertical distance compared to magnetic energy
as well as towards  the axis; this picture depends on the background
local disk-parameters (with larger $\beta $ the area of jet with
$M_{zA}(r,z) < 10^4$ gets wider).

Considered weakly magnetized accretion disk-relativistic jet solutions
describe the formation of the high velocity magnetized jet from slowly
accreting locally Keplerian electron-ion-photon disk-flow and can be used to
analyze topological characteristics of Jets from binary systems linking them
to the local physical conditions at the jet launching areas; since in the
present study the clear link of outflow parameters to the background
accretion disk conditions is well shown the predictions for observed
structures are possible. The further kinematic acceleration and collimation
of the jet flow may happen due to the effects not considered in the
current minimal model.

\section*{Acknowledgments}
\label{Ackno}

Authors express special thanks to V.I. Berezhiani and A.G. Tevzadze
for their valuable discussions. Present research was partially
supported by Shota Rustaveli Georgian National Foundation Grant Project No.
FR-22-8273.



\appendix
\section{Derivation of major analytical relations}

Following \cite{SY2011,Yoshida2012} we introduce a so-called ``ideal''
and ``reduced'' factors of ``local'' density (see also Arshilava et al 2019)
to seek for the steady state solutions of the disk-jet structures persisting
around a central accreting slowly rotating compact object:
\begin{equation}
\rho = \rho_{I} \rho_{R} ~
\label{SS15}
\end{equation}
to separate the ideal fluid and the dissipative effects and to track
the accretion effects in disk-jet structure formation process so that
$\rho_{R} = 1$ and $\rho_{I} = \rho$ in conventional ideal fluid mechanics
with zero dissipation. The  {``ideal''} and ``reduced'' momenta then read as:
\begin{equation}
{\bf P}_{I} = \rho_{I} {\bf v} ~, \qquad \qquad {\bf P}_{R} = \rho_{R} {\bf v}
\label{SS16}
\end{equation}
and from the equations (\ref{PDE1},\ref{PDE2}) we obtain a ``Generalized Pressure
Balance equation'' in terms of the new momenta:
\begin{equation}
{\bf P}_{R} \times (\nabla \times {\bf P}_{R}) = \label{Bernoulli1}
\hskip 4 cm
\end{equation}
\begin{equation*}
= {1 \over 2} \nabla P_{R}^2 + \rho_{R}^2 \nabla \left(h + \Phi \right)
+ \frac{\rho_{R}}{\rho_{I}} \left[ {\bf P}_{R}
(\nabla \cdot {\bf P}_{I}) + \nabla \cdot {\rm T} \right] .
\end{equation*}
Due to the geometry of the problem the natural assumption for the reduced
momentum to obey the Beltrami Condition implying that the Reduced Momentum
is aligning the corresponding Generalized Vorticity is valid
(see e.g. \cite{yso} and references therein):
\begin{equation}
{\bf P}_R = \lambda \ ( \nabla \times {\bf P}_R ) ~, \label{Beltrami1}
\end{equation}
making the l.h.s. term of the Eq. (\ref{Bernoulli1}) strictly zero.
Now $\lambda $ stands for the Beltrami parameter related
to the composite flow "reduced momentum" for which we seek the solution
such that it makes the last term zero on the r.h.s. of
the Eq. (\ref{Bernoulli1}) being fully determined by the accretion
(i.e. total viscosity effect dominated by photon gas):
\begin{equation}
{\bf P}_R (\nabla \cdot {\bf P}_I) + \nabla \cdot {\rm T} = 0 ~.
\label{Bernoulli21}
\end{equation}
Then, in such minimal model the pressure balance equation (\ref{Bernoulli1})
reduces to a so-called ``Generalized Bernoulli Condition''
\begin{equation}
{1 \over 2} \nabla P_R^2 + \rho_R^2 \nabla \left(h + \Phi\right) = 0 ~.
\label{Bernoulli22}
\end{equation}
Hence, the stationary state of the rotating thin composite system of AD
can be fully investigated using Eqs. (\ref{SS13},\ref{Beltrami1},\ref{Bernoulli21},
\ref{Bernoulli22}) and the explicit form of the viscous stress tensor related
to the specific object conditions. Below we employ the model in which
the small scale turbulence creates the anomalous dissipation that can be
described by using the generalized $\alpha$--viscosity model introduced by \cite{Shakura1973}
in which we use the effective viscosity model both in the disk and
jet as well as in the disk-jet transition areas (\citealt{yso}).
In cylindrical coordinates \ $(r, \varphi, z)$ \ describing the
disk-jet system, keeping the only significant components $T_{r \varphi}$ and
$T_{z \varphi}$ of the viscous stress tensor $T_{ik}$, assuming the
strong azimuthal rotation, splitting the pressure $\cal{P} $ into the
background constant component ${\cal P}_0$ and deviation form it $p$ ,
\begin{equation}
{\cal{P}} = {\cal P}_0 + p \ ,
\end{equation}
the turbulent stress tensor can be split into the background
constant component $\bar T_{ik}$ and smaller varying deviation $t_{ik}$:
\begin{equation}
T_{ik} = \bar T_{ik} + t_{ik}
\label{Tik}
\end{equation}
with $\bar T_{r \varphi} = \alpha_0 {\cal P}_0 $. Following \cite{yso},
assuming axisymmetric flow $v_\varphi = r \Omega_{\rm K}(r,z)$ rotating
locally with Keplerian angular velocity (justified for the rotationally
supported flow, for which the radial pressure gradients can be negligible compared to
the centrifugal force -- situation found for slowly accreting flows
with slowly varying background pressure ${\cal P}_0 \simeq const $ ):
\begin{equation}
\Omega_{\rm K}^2(r,z) = {G M_\star \over (r^2 + z^2)^{3/2}} ~, \label{Omega_Kep}
\end{equation}
where $M_\star$ is the mass of the central object, in the axisymmetric
case, the viscous stress tensor elements can be calculated as follows:
\begin{equation}
t_{r \varphi} = {r^2 \over r^2 + z^2} \,\beta \,p ~, \label{Tr}
\end{equation}
\begin{equation}
t_{z \varphi} = {r z \over r^2 + z^2}\, \beta  \,p ~, \label{Tz}
\end{equation}
where positive/negative $p$ corresponds to the stronger/weaker
turbulence compared to the background generalized turbulent steady state.

\bigskip

The geometry of the observed disk-jet structures and the continuity
Eq. (\ref{PDE2}) dictate the expansion of the flow velocity
in the following way (\citealt{SY2011,yso}):
\begin{equation}
{\bf v} = {1 \over \rho} \left( \nabla \psi \times \nabla \varphi \right)
+ r v_\varphi \nabla \varphi ~,
\label{Stream}
\end{equation}
where the axisummetric stream function $\psi$ exactly matches the stream
function of the actual momentum ${\bf P} = \rho {\bf v}$ of the bulk flow.
Then, the Eq. (\ref{Bernoulli21}) takes the form:
\begin{equation}
{v_\varphi \over r} \left( {\partial \psi \over
\partial r} {\partial \over
\partial z} \ln \rho_R - {\partial \psi \over \partial z} {\partial \over
\partial r} \ln \rho_R
 \right)
 \label{Stream1}
\end{equation}
\begin{equation*}
\hskip 2cm
= \beta \,{r^2  \over r^2 + z^2} \left[ {\partial p \over \partial r}
+ {z \over r} {\partial p \over \partial z} + {2\beta \over r} p \right] ~,
\end{equation*}
and the generalized Bernoulli Condition (\ref{Bernoulli22}) reduces to:
\begin{equation}
\nabla {\cal E}_m + \left(v_\varphi^2 + {(\nabla \psi)^2 \over r^2 \rho^2} \right)
\nabla \ln \rho_R = 0 ~, \label{Stream2}
\end{equation}
where we have introduced the total mechanical energy
\begin{equation}
{\cal E}_m \equiv \Phi + {v_\varphi^2 \over 2} +
{(\nabla \psi)^2 \over 2 r^2 \rho^2} ~. \label{E}
\end{equation}
The system of equations describing the relativistic disk-powerful jet
equilibrium structure can be closed using the Beltrami condition
(\ref{Beltrami1}) rewritten using the stream function:
\[
\nabla\times\nabla\psi\times\nabla\varphi \ + \ \nabla\times
\left[(\rho r v_{\varphi}) \nabla\varphi \right]
\ + \ \nabla\psi\times\nabla\varphi\times\nabla\ln\rho_I
\]
\begin{equation}
+ \ (\rho r v_{\varphi}) \nabla\varphi\times\nabla\ln\rho_I =
\lambda\left(\nabla\psi\times\nabla\varphi
+ (\rho r v_{\varphi}) \nabla\varphi\right) \ .
\label{beltrami2}
\end{equation}
Hence, Eqs. (\ref{Stream1},\ref{Stream2},\ref{E}) and (\ref{beltrami2})
constitute the final set of equations.

\bigskip

To construct the solutions representing disk-jet structure,
following \cite{SY2011}, we introduce the orthogonal variables
$\tau = z/r$ and $\sigma = \sqrt{r^2 + z^2}$ so that
$\nabla \tau \cdot \nabla \sigma = 0$ leading to the obvious
results for the Gravitational Potential to be $\Phi=\Phi(\sigma)
= - \, \Omega_0^2/\sigma $ and the stream function to be dependent
only on the $\tau$ variable \ ($\psi = \psi(\tau) $):
thus, making it possible to separate the variables in the solution.
Then, the Eq. (\ref{Stream1}) will take the following form in the
similarity variables,
\begin{equation}
v_\varphi {1 + \tau^2 \over \sigma^2} {\partial \psi \over \partial \tau}
{\partial \over \partial \sigma} \ln \rho_R  = \beta \,
\left( {\partial \over \partial \sigma} + {2 \over \sigma} \right)
{p \over 1+\tau^2} ~.
\label{Sim1}
\end{equation}
The three components $r, \varphi, z$ of the Beltrami conditions
(\ref{beltrami2}), after some straightforward algebra, in the new variables
yield
\begin{equation}
{\partial \over \partial \sigma} \left( {\sigma \over (1+\tau^2)^{1/2}}
\rho_R v_\varphi \right) =  0 ~,
\label{Sim2}
\end{equation}

\[
{\partial^2 \psi \over \partial \tau^2} + \left( {3 \tau \over 1+ \tau^2} -
{\partial \over \partial \tau} \ln \rho_I \right)
{\partial \psi \over \partial \tau}
\]
\begin{equation}
= - \lambda \ {\sigma^3 \over (1+\tau^2)^{5/2}} \rho v_\varphi ~,
\label{Sim4}
\end{equation}

\begin{equation}
{\partial \over \partial \tau} \ln \left( {\sigma \over (1+\tau^2)^{1/2}}
\rho_R v_\varphi \right) =
\lambda \ {(1+\tau^2)^{1/2} \over \sigma \rho v_\varphi}
{\partial \psi \over \partial \tau}  ~,
\label{Sim3}
\end{equation}
while $r$ and $z$ components of the Generalized Bernoulli Condition
(\ref{Bernoulli22}) lead to the following equations, respectively:
\begin{equation}
{1 \over \rho} {\partial p \over \partial \sigma} +
{\partial {\cal E}_m \over \partial \sigma} + 2\left( {\cal E}_m - \Phi \right)
{\partial \over \partial \sigma} \ln \rho_R = 0 ~, \label{Sim5}
\end{equation}
\begin{equation}
{1 \over \rho} {\partial p \over \partial \tau} +
{\partial {\cal E}_m \over \partial \tau} + 2\left( {\cal E}_m - \Phi \right)
{\partial \over \partial \tau} \ln \rho_R = 0 ~, \label{Sim6}
\end{equation}
with the total mechanical energy of the composite system ${\cal E}_m=
{\cal E}_m (\sigma, \tau )$. Hence, in similarity variables the physical
system of our problem can be analyzed using Eqs. (\ref{Sim1} - \ref{Sim6}).

\bigskip

We seek the solution of the system assuming that describing physical variables
($p, \rho_{\rm I}, \rho_{\rm R}, \lambda $) can be factorized
using the similarity variables $(\sigma, \tau $)
applying variable splitting ansatz (like e.g. $p(\sigma,\tau) =
p_1(\sigma) p_2(\tau) $, see (\citealt{SY2011,yso})).
For the azimuthal velocity we have:
\begin{equation}
v_\varphi(\sigma,\tau) = V_{\rm Kep} = {\sigma_0 \Omega_0 \over (1+\tau^2)^{1/2}}
\left( {\sigma \over {\sigma_0}}\right)^{-1/2} ~, \label{Kepler}
\end{equation}
where $\Omega_0$ is the Keplerian angular velocity of the rotation at some
characteristic radius $\sigma_0$ in the disk. Applying this ansatz into Eqs.
(\ref{Sim1}-\ref{Sim6}), solving them in $\sigma$,  we find that the balance of
all terms give the explicit solutions for all ($p, \rho_{\rm I}, \rho_{\rm R},
\lambda $) \ in the $\sigma$ coordinate. After finding the radial profiles
of these parameters above equations represent the ordinary differential
equations with respect to \ $\tau$--variable from which, after
straightforward algebra, one obtains (below \ $\rho_2 (\tau )
\equiv \rho_{I2} \rho_{R2}$ \ ):
\begin{equation}
\beta p_2 = \Omega_0\,\sigma_0^{3/2}\,(1+\tau^2)^{3/2} {{\rm d} \psi \over {\rm d} \tau}  ~,
\end{equation}
\begin{equation}
p_2 \rho_2 + { 2 \over 5} (1+\tau^2)^{3}
\left({{\rm d} \psi \over {\rm d} \tau}\right)^2
= { 2 \over 5} \, \Omega_0^2 \,\sigma_0^{3} \frac{{\tau^2}}{1 + \tau^2}
\rho_2^2  ~,
\end{equation}
from which one derives the defining algebraic equation
\begin{equation}
\beta \left[ \beta W(\tau)^2 +
{5 \over 2} { \sigma_0^{3/2} \Omega_0 \over (1+\tau^2)^{3/2}} W(\tau) -
\beta \frac{\sigma_0^3 \Omega_0^2 \tau^2}{(1+\tau^2)^4} \right]  =  0
\label{Realizability}
\end{equation}
for the function
$$ W(\tau ) \equiv  {1 \over \rho_2} {{\rm d} \psi \over {\rm
d} \tau} \ .
$$

As shown in \cite{yso}, the Eq. (\ref{Realizability}) links the values of
$\beta$ parameter (defining the generalized viscosity of our composite
3-fluid system), $\tau$-dependent part of density $\rho_2$ and the stream-function $\psi$,
and represents the "realizability" condition for all existing solutions
within the considered Beltrami flow model of disk-jet structure formation.
There were found three apparent solutions to this equation: 1) one
with $\beta=0$, corresponding to the background dissipation model
($T_{ik} = \alpha_0 {\cal P}_0$ ) and 2) two for $\beta \neq 0$
which for small $\beta \ll 1$ can be approximated by $W_+(\tau)$
(corresponds to disk-solution) and $W_-(\tau)$ (corresponds to the jet one),
where
\begin{equation}
W_+(\tau) \approx {2 \over 5} \ \beta \sigma_0^{3/2} \Omega_0
{\tau^2 \over (1+\tau^2)^{5/2}} > 0 ~, \label{Wplus}
\end{equation}
\begin{equation}
W_-(\tau) \approx -{5 \over 2} \,{\sigma_0^{3/2} \Omega_0 \over \beta}
{1\over (1+\tau^2)^{3/2}} < 0 ~,
\label{Wminus}
\end{equation}
making it possible to find the dependence of the Beltrami
parameter $\lambda$ on the turbulent viscosity parameter
$\beta$ from the Eqs. (\ref{Sim1} - \ref{Sim6}); moreover,
derived two separate solutions grow with $\beta$
($ \lambda_{\pm} = \lambda(W_{\pm}) \propto \beta  $).
Illustration for $W(\tau)$ for the three region solution
-- disk-jet configuration in one solution describing the classes
of solutions corresponding to the accretion disk–-ejection
jet flow (one flow with the matter and total energy) can be found
in \cite{yso} (see its Figure 2 and related discussion on Beltrami-flow
model for disk-jet structure formation). In this model a non-uniformly
($\beta \neq 0$) viscous accreting flow in the disk region
($\tau < \tau_d$) goes through uniformly viscous ($\alpha_0 =const,
\ \beta =0$) ballistic regime in the transition region
($\tau_d < \tau < \tau_j$) and, finally, into the non-uniformly
viscous outflow configuration in the jet region ($\tau > \tau_j$)
advecting/dragging the magnetic field.

The general solutions of our weakly magnetized
disk-jet model  can be calculated as follows (with the general
solution $W(\tau)$  of Eq. (\ref{Realizability})
in which $\beta $ represents the contributions from
generalized turbulent viscosity \citep{Shakura1973}):
\begin{eqnarray}
\rho(\sigma,\tau) &=& \sigma^{-3/2} \rho_2(\tau)
\label{Solrho} ~ ,\\
p(\sigma,\tau) &=& {\sigma_0^{3/2} \Omega_0 \over \beta}
{(1+\tau^2)^{3/2} \over \sigma^{5/2}} \rho_2(\tau) W(\tau) ~ ,
\label{Sol1} \\
v_r(\sigma,\tau) &=& - \, {1+\tau^2 \over \sigma^{1/2}} W(\tau) ~ ,
\label{SolVr} \\
v_z(\sigma,\tau) &=& - \, \tau \,{1+\tau^2 \over \sigma^{1/2}} W(\tau) ~ .
\label{Sol2}
\end{eqnarray}
Eqs. (\ref{Solrho}-\ref{Sol2}) together with the radial profiles
and appropriate choice of the solution for $W$ give the full
solution for the disk-jet bulk (ion and electron-photon)
flow for different types of density profiles $\rho_2(\tau)$ for
which we employ the power-law distribution (cf. \citealt{SY2011}):
\begin{equation}
\rho_2(\tau) = A_{\rm d} (\tau + \tau_0)^{m_{\rm d}} +
A_{\rm j} (\tau + \tau_0)^{m_{\rm j}} ~,
\label{Rho2SY}
\end{equation}
where parameters $A_d$, $m_d$ and $A_j$, $m_j$ define the density profile in
the disk and jet regions, respectively. The small parameter $\tau_0 \ll 1$ is
used to avoid divergence at the disk center ($\tau=0$). Hence, Eqs.
(\ref{Solrho}-\ref{Sol2}) and ({\ref{Rho2SY}}) describe classes of disk-jet
solutions within our self-similar Beltrami-flow model.

Substitution of Eqs. (\ref{Wplus},\ref{Wminus}) into (\ref{SolVr}) and
(\ref{Sol2}) yields the velocity field components of the
weakly magnetized disk flow.


\begin{thebibliography}{00}



\bibitem[ ()]{}
\bibitem[Arce et al. (2013)]{Acre2013}
    Arce, H. G., Mardones, D., Corder, S. A.,
    Garay, G., Noriega-Crespo, A.,
    Raga, A. C., 2013
    Aastrophys.J., 774, 39

\bibitem[Arshilava et al. (2019)]{yso}
    Arshilava, E. Gogilashvili, M., Loladze, V., Jokhadze, I.,
    Modrekiladze, B., Shatashvili, N.L. Tevzadze, A.G. 2019,
    J. High Energy Astrophysics {\bf 23}, 6

\bibitem[Bai (2016)]{Bai}
    Bai, X-N., 2016
    Astrophys. J., 821, 80

\bibitem[Balbus \& Hawley (1991)]{BH91}
    Balbus, S.A. \& Hawley, J.F.  1991
    Astrophys. J., 376, 214

\bibitem[Bally (2016)]{Bally2016}
    Bally, J., 2016
    Ann. Rev. Astron. Astrophys., 54, 491

\bibitem[Balman (2020)]{Balman}
      Balman S. 2020,
      ASR {\bf 66}, 5, 1097

\bibitem[Barnier et al. (2023)]{Ferreira23}
    Barnier S., Petrucci, P.-O., Ferreira, J., Marcel, G. 2023
    Astronomische Nachrichten, 344(4) 20230020

\bibitem[Baierlein (1978)]{Baierlein}
    Baierlein, R., 1978
    MNRAS {\bf 184}, 843

\bibitem[Barnaveli \& Shatashvili (2017)]{BS-flow}
    Barnaveli, A.A., Shatashvili, N.L. 2017
    Astrophys Space Sci. {\bf 362}, 164

\bibitem[Begelman et al (1984)]{Begelman}
    Begelman, M.C., Blandford, R.D., and
    Rees, M.D. 1984
    Rev. Mod. Phys. {\bf 56} 255

\bibitem[Begelman (1993)]{begelman4}
    Begelman, M. C., 1993
    ''Conference summary'', in {\it Astrophysical
    Jets}, \ ed. D. Burgarella et al (Cambridge: Cambridge Univ.
    Press), 1993, pp. 305-315.

\bibitem[Begelman (1998)]{begelman5}
    Begelman, M. C., 1998
    Aastrophys. J., 493, 291

\bibitem[Berezhiani et al (2015)]{BSM_deg}
    Berezhiani, V.I., Shatashvili, N.L.
    and Mahajan, S.M. 2019
    Phys. Plasmas {\bf 22}, 022902

\bibitem[Bisnovatyi-Kogan \& Lovelace (2007)]{lovelace2}
    Bisnovatyi-Kogan G. S. and Lovelace, R.V.E., 2007,
    ApJ 667(2), L167-L169

\bibitem[Blandford \& Rees (1974)]{bland}
   Blandford,  R. D. and Rees, M. J., 1974
   MNRAS, 169, 395

\bibitem[Blandford \& Znajek (1997)]{bland1}
    Blandford, R. D. and Znajek, R. L., 1977
    MNRAS, 179, 433

\bibitem[Blandford \& Payne (1982)]{bland3}
    Blandford, R. D. and Payne, D. G., 1982
    MNRAS, 199, 883

\bibitem[Blandford (1994)]{bland2}
    Blandford, R. D., 1994
    ApJS, 90, 515

\bibitem[Celloti \& Blandford (2001)]{bland4}
    Celotti, A.  and Blandford, R. D., ''Black Holes in Binaries and
    Galactic Nuclei: Diagnostics, Demography and Formation'', in {\it
    ESO Astrophysics Symposia} \  ed. L. Kaper \textit{et al.}
    (Berlin, Heidelberg: Springer-Verlag), 2001, 206

\bibitem[Chan \& Jones (1975)]{Chan}
    Chan, K. L., \& Jones, B. J. T. 1975,
    Astrophys. J., 200, 454

\bibitem[Clarke \& Alexander (2016)]{Clarke2016}
    Clarke, C. J., and Alexander, R. D., 2016
    MNRAS, 460, 3044

\bibitem[Del Santo et al. (2023)]{FerreiraUltra}
    Del Santo, M., Pinto, C., Marino, A.,
    D’Aı, A.,Petrucci, P.-O., Malzac, J., Ferreira, J.
    Pintore, F. Motta, S. E., Russell, T.D.,  Segreto, A.
    and Sanna, A. 2023
    MNRAS 523, L15–L20

\bibitem[Dullemond et al. (2007)]{Dullemond07}
    Dullemond, C. P., Hollenbach, D., Kamp, I., \&
    D’Alessio, P. 2007,
    in Protostars and Planets V, ed. B. Reipurth,
    D. Jewitt, \& K. Keil (Tucson, AZ:
    Univ. Arizona Press), 555

\bibitem[Ferreira (1997)]{Ferreira1997}
    Ferreira, J., 1997
    A\&A 319, 340

\bibitem[Ferreira et al. (2006)]{Ferreira}
    Ferreira, J., Dougados, C. and Cabrit, S., 2006
    A\&A 453, 785

\bibitem[Ferreira et al. (2007)]{Ferreira07}
    Ferreira, J., Dougados, C. and Whelan, E., 2007
    ”Jets from Young Stars I: Models and Constraints” in Lecture Notes in Physics
    ed. Ferreira, J.  et al. (Berlin, Heidelberg: Springer-Verlag) 2007, 723, 181.

\bibitem[Ferreira (2008)]{Ferreira08}
    Ferreura, J. 2008
    New Astronomy Reviews 52, 42–59

\bibitem[Harrison (1973)]{Harrison1973}
    Harrison, E.R. 1973
    MNRAS 165, 185

\bibitem[Hartmann (2009)]{Hartmann}
    Hartmann, L., 2009
    Accretion Processes in Star Formation,
    Cambridge University Press

\bibitem[Ioannidis \& Froebrich (2012)]{Ioannidis2012}
    Ioannidis, G., Froebrich, D., 2012,
    MNRAS, 421, 3257

\bibitem[Kafka \& Honeycutt (2004)]{kafka}
    Kafka, S. \& Honeycutt, R.K. 2004
    Astrophys. J. {\bf 128}, 2420

\bibitem[Kawka \& Vennes (2014)]{DAZ}
    Kawka, A. and Vennes, S. 2014
    MNRAS {\bf 439}, L90

\bibitem[Kepler et al (2013)]{kepler}
    Kepler, S. O., Pelisoli, I.,
    Jordan, S., Kleinman, S.J., Koester, D., K\"ulebi, D.B.,
    Pecanha, B.V., Castanheira, B.G., Nitta, A., Costa, J.E.S.,
    Winget, D.E., Kanaan, A. and Fraga, L. 2013 
    MNRAS {\bf 429}, 2934

\bibitem[Koide et al. (1998)]{Koide}
    Koide, S., Shibata, K. and Kudoh, T., 1998
    Astrophys. J., 495, L63

\bibitem[Kotorashvili et al (2020)]{RD_deg}
    Kotorashvili, K., Revazashvili, N., Shatashvili, N.L. 2020
    Astrophys. Space Sci., {\bf 365}, 175

\bibitem[Kotorashvili \& Shatashvili (2022)]{RD_2T}
    Kotorashvili, K. and Shatashvili, N.L. 2022
    Astrophys. Space Sci., {\bf 367}, 2

\bibitem[Krasnopolsky et al. (1999)]{bland3}
    Krasnopolsky, R., Li, Z. Y.  and Blandford, R. D., 1999
    Aastrophys.J.,, 526, 631

\bibitem[Kudoh \& Shibata (1997)]{shibata}
    Kudoh, T. and Shibata, K., 1987
    Aastrophys.J.,, 474, 362

\bibitem[Kuwabara et al. (2005)]{shibata2}
    Kuwabara, T., Shibata, K., Kudoh,  T. and Matsumoto, R., 2005
    ApJ, 621, 921

\bibitem[Lee et al. (2018)]{Lee2018}
    Lee, C.-F., Li, Z.-Y., Codella, C.,
    Ho, P. T. P., Podio, L., Hirano, N.,
    Shang, H., Turner, N. J., Zhang, Q., 2018
    ApJ, 856, 14

\bibitem[Livio (1997)]{Livio}
    Livio, M. ”The Formation Of Astrophysical Jets",
    in Accretion Phenomena and Related Outflows;
    IAU Colloquium 163 ed. D. T. Wickramasinghe et al
    (San Francisco: ASP) ASP Conference Series 1997, 121, 845.

\bibitem[Long et al (2002)]{AD_modeling}
    Long, K.S., Knigge, C. 2002
        Astrophys. J. {\bf 579}, 725

\bibitem[Lovelace et al. (1994)]{lovelace1}
    Lovelace, R.V.E., Romanova, M.M. and Newman, W.I., 1994,
    ApJ, 437, 136

\bibitem[Lovelace (1976)]{Lovelace3}
    Lovelace, R. V. E. 1976,
    Nature, 262, 649

\bibitem[Mahajan et al. (2002)]{mnsy}
    Mahajan S. M., Nikol'skaya K. I.,
    Shatashvili N. L., Yoshida Z., 2002
    Astrophys. J., 576, L161

\bibitem[Mahajan et al (2001)]{mmns-1}
    Mahajan, S.M., Miklaszewski, R.,
    Nikol’skaya, K.I. and Shatashvili, N.L. 2001
    Phys. Plasmas {\bf 8}, 1340

\bibitem[Mahajan et al (2005)]{msms1}
    Mahajan, S.M., Shatashvili, N.L.,
    Mikeladze, S.V. and Sigua, K.I. 2005
    Astrophys. J. {\bf 634}, 419

\bibitem[Mahajan et al (2006)]{msms2}
    Mahajan, S.M., Shatashvili, N.L.,
    Mikeladze, S.V. and Sigua, K.I. 2006 
    Phys. Plasmas {\bf 13}, 062902

\bibitem[Matsumoto \& Tajima (1995)]{Matsumoto1}
    Matsumoto, R. and Tajima, T. 1995
    Aastrophys.J., 445, 767

\bibitem[Matsumoto et al. (2004)]{Matsumoto2}
    Matsumoto, r., Machida, M. and Nakamura, K. 2004
    Prog. Theor. Phys. Suppl. 155, 124

\bibitem[Mignone et al (2005)]{mignone}
    Mignone, A., Plewa, T. \& Bodo, G. 2005
    Astrophy. J. Supp. {\bf 160}, 199

\bibitem[Meier et al. (1994)]{Meier}
    Meier, D. L., Edgingon, S., Godon, P.,
    Payne, D. G., \& Lind, K. R. 1997,
    Nature, 388, 350

\bibitem[Mirabel \& Rodriguez (1994)]{Mirabel94}
    Mirabel, I. F., \& Rodriguez, L. F. 1994,
    Nature, 371, 46
   
\bibitem[Mirabel \& Rodriguez (1999)]{Mirabel}
    Mirabel, I.F. \& Rodriguez, L.F., 1999
    Annu. Rev. Astron. Astrophys.  37:409–43

\bibitem[Mouchet et al (2012)]{Mouchet}
    Mouchet, M., Bonnet-Bidaud, J.M., de Martino, D. 2012
    Mem. SAI {\bf 83}, 578

\bibitem[Mukai (2017)]{mukai}
    Mukai, K. 2017
    {\it PASP} {\bf 129}, 062001

\bibitem[Nordhaus et al (2010)]{Disk-DB}
    Nordhaus, J., Wellons, S., Spiegel, D.S.,
    Metzger, B.D., Blackman, E.G., 2010
    Formation of high-field magnetic white dwarfs
    from common envelopes, Proceedings of the National Academy of Sciences,
    {\bf 108(8)}, 3135

\bibitem[O'Dell (1981)]{odell}
    O'Dell, 1981
    Astrophys. J. 243, L147

\bibitem[Ohsaki et al (2001)]{osym1}
    Ohsaki, S., Shatashvili, N.L.,
    Yoshida, Z. and Mahajan, S.M. 2001 
    Astrophys. J. {\bf 559}, L61; \


\bibitem[Ohsaki et al (2002)]{osym2}
    Ohsaki, S., Shatashvili, N.L.,
    Yoshida, Z. and Mahajan, S.M. 2002 
    Astrophys. J. ASTROPHYS. J., {\bf 570}, 395

\bibitem[Ouyed et al. (1997)]{Pudritz2}
    Ouyed, R., Pudritz, R. E., \&
    Stone, J. M. 1997,
    Nature, 385, 409

\bibitem[Peebles (1969)]{Peebles1}
    Peebles, P. J. E., 1969,
    Astrophys. J. 155, 393.

\bibitem[Peebles (1980)]{Peebles2}
    Peebles, P. J. E., 1980,
    The Large-Scale Structure of the Universe
    (Princeton University, Princeton, NJ).

\bibitem[Peebles (1993)]{Peebles3}
    Peebles, P. J. E., 1993,
    Principles of Physical Cosmology (Princeton University, Princeton, NJ).

\bibitem[Pelletier et al. (1971)]{Pelletier}
    Pelletier, G., Ferreira, J., Henri, G.,
    \& Marcowith, A. 1996, in Solar and
    Astrophysical Magnetohydrodynamic Flows, ed. K. C.
    Tsinganos (Dordrecht: Kluwer), 643

\bibitem[Phiney (1982)]{Phiney}
    Phinney, E.S., 1982
    MNRAS 198, 1109

\bibitem[Pudritz \& Norman (1986)]{Pudritz}
    Pudritz, R. E., \& Norman, C. 1986,
    Aastrophys.J., 301, 571

\bibitem[Puebla et al (2011)]{Diaz}
    Puebla, R.E., Diaz, M.P., Hillier, D.J., Hubeny, I. 2011
    Astrophys. J., {\bf 736}, 17

\bibitem[Shakura \& Sunyaev (1973)]{Shakura1973}
    Shakura, N. I., Sunyaev, R. A., 1973,
    A\&A, 24, 337

\bibitem[Shatashvili \& Yoshida (2011)]{SY2011}
    Shatashvili, N.L. and Yoshida, Z., 2011,
    AIPCP, 1445, 34-53

\bibitem[Shibata \& Uchida (1986)]{shibata3}
    Shibata, K., \& Uchida, Y. 1986,
    PASJ, 38, 631

\bibitem[Smith et al. (2014)]{Smith2014}
    Smith, M. D., Davis, C. J., Rowles, J. H., Knight, M., 2014,
    MNRAS, 443, 2612

\bibitem[Takahara et al. (1989)]{takahara}
    Takahara, F., Rosner, R., \& Kusunose, M. 1989,
    Aastrophys.J., 346, 122

\bibitem[Takasao et al. (2017)]{shibata4}
    Takasao, S., Suzuki1, T.K., and Shibata, K. 2017
    Aastrophys.J., 847, 46

\bibitem[Tremblay et al (2015)]{tremblay}
    Tremblay, P.-E., Fontaine, G., Freytag, B.,
    Steiner, O., Ludwig, H.-G., Steffen, M., Wedemeyer, S.
    and Brassard, P. 2015
    Astrophys. J., {\bf 812}, 19

\bibitem[Tout et al (2008)]{Ferrario-3}
    Tout C.A, Wickramasinghe D.T., Liebert J,
    Ferrario L. and Pringle J.E. 2008
    MNRAS, {\bf 387}, 897

\bibitem[Sikora \& Wilson (1981)]{Sikora}
    Sikora, M., \& Wilson, D.B., 1981
    MNRAS 197, 529

\bibitem[Spruit (2010)]{Spruit1}
    Spruit, H. C. 2010
    The Jet Paradigm, Lecture Notes in Physics, 794.
    Springer-Verlag Berlin Heidelberg, 2010, p. 233

\bibitem[Spruit (2011)]{Spruit2}
    Spruit, H.C. 2011
    AIP Conference Proceedings, 1381, 227

\bibitem[Uchida \& Shibata (1985)]{uchida}
    Uchida, Y., \& Shibata, K. 1985,
    PASJ, 37, 515

\bibitem[Ustyugova et al. (1999)]{Ustyugova}
    Ustyogova, G.V., Koldova, A.V., Romanova, M.M.,
    Chechetkin, V.M., Lovelace, R.V.E., 1999
    Astrophys. J. 516, 221

\bibitem[Yang et al. (2024)]{yuan}
    Yang, H., Yuan, F., Li, H., Mizuno, Y., Guo, F.,Lu, R.,Ho, L.C.,
    Lin, X. Zdziarski, A.A., Wang, J., 2024
    Sci. Adv. 10, eadn3544

\bibitem[Yoshida \& Shatashvili (2012)]{Yoshida2012}
    Yoshida, Z., Shatashvili N. L., 2012,
    arXiv:1210.3558

\bibitem[Vlemmings et al. (2007)]{Vlemmings}
    Vlemmings, W. H. T., Bignall, H. E.
    and Diamond,P. J. 2007
    ApJ 656, 198

\bibitem[Zanni et al. (2007)]{zanni}
    Zanni, C., Ferrari, A., Rosner, R., Bodo,  G. and Massaglia, S., 2007
    A\&A, 469, 811

\bibitem[Zhang et al. (2014)]{Zhang2014}
    Zhang, M., Wang, H., Hennings, T., 2014,
    AJ, 148, 26

\bibitem[Weinberg (1972)]{Weinberg}
    Weinberg, S., 1972
    Gravitation \& Cosmology, J. Wiley, New York

\bibitem[Widrow (2002)]{Widrow}
    Widrow, L., 2002
    Rev. Mod. Phys. {\bf 74}, 1

\bibitem[Winget \& Kepler (2008)]{winget}
    Winget, D.E.  and Kepler, S.O. 2008
    Annu. Rev. A\&A {\bf 46}, 157

                                                                                            %


\end{thebibliography}



\end{document}